\documentclass[1p,times,letterpaper,aas_macros]{elsarticle}
\usepackage{lipsum}
\usepackage{amsmath}
\usepackage{mathtools}
\usepackage{textcomp}
\usepackage[hyperfootnotes=false]{hyperref}
\usepackage[table]{xcolor}
\hypersetup{linkbordercolor=green}
\usepackage{caption}
\usepackage{adjustbox}
\usepackage{upgreek}
\usepackage{pifont}
\newcommand{\cmark}{\ding{51}}%
\newcommand{\xmark}{\ding{55}}%
\usepackage{color,soul}
\usepackage{float}
\floatstyle{plaintop}
\restylefloat{table}
\usepackage[bottom]{footmisc}
\usepackage{enumitem}

\journal{Nuclear Instruments and Methods in Physics Research A}

\usepackage{lineno}
\graphicspath{{FinalFigures/}}

\title{Passivation of Si(Li) detectors operated above cryogenic temperatures for space-based applications}

\author{N.~Saffold\textsuperscript{a,}\footnote{Corresponding author\newline \textit{Email:} \href{mailto:nas2173@columbia.edu}{{nas2173@columbia}}},
	F.~Rogers\textsuperscript{b},
	M.~Xiao\textsuperscript{b},
	R.~Bhatt\textsuperscript{b},
	T.~Erjavec\textsuperscript{b},
	H.~Fuke\textsuperscript{c},	
	C.~J.~Hailey\textsuperscript{a},\\
	M.~Kozai\textsuperscript{c},
	D.~Kraych\textsuperscript{a},
	E.~Martinez\textsuperscript{a},
	C.~Melo-Carrillo\textsuperscript{a},
	K.~Perez\textsuperscript{b},
	C.~Rodriguez\textsuperscript{a},\\
	Y.~Shimizu\textsuperscript{d},
	B.~Smallshaw\textsuperscript{a}}

\address{\textsuperscript{a}Columbia Astrophysics Laboratory, Columbia University, New York, NY 10027\\\textsuperscript{b}Department of Physics, Massachusetts Institute of Technology, Cambridge, MA 02139\\ \textsuperscript{c}Institute of Space and Astronautical Science, Japan Aerospace Exploration Agency (ISAS/JAXA), Sagamihara, Kanagawa 252-5210, Japan\\ \textsuperscript{d}Kanagawa University, Yokohama, Kanagawa 221-8686, Japan}

\ead{nas2173@columbia.edu}

\begin{document}
\begin{abstract}
This work evaluates the viability of polyimide and parylene-C for passivation of lithium-drifted silicon (Si(Li)) detectors. The passivated Si(Li) detectors will form the particle tracker and X-ray detector of the General Antiparticle Spectrometer (GAPS) experiment, a balloon-borne experiment optimized to detect cosmic antideuterons produced in dark matter annihilations or decays. Successful passivation coatings were achieved by thermally curing polyimides, and the optimized coatings form an excellent barrier against humidity and organic contamination. The passivated Si(Li) detectors deliver $\lesssim$\,4\,keV energy resolution (FWHM) for 20$-$100\,keV X-rays while operating at temperatures of $-$35 to $-$45\,\textdegree{C}. This is the first reported successful passivation of Si(Li)-based X-ray detectors operated above cryogenic temperatures.
\end{abstract}

\maketitle

\section{Introduction\label{s-intro}}
A key technical challenge in lithium-drifted silicon (Si(Li)) detector fabrication is passivation to protect against environmental contaminants. The General Antiparticle Spectrometer (GAPS) experiment has previously reported on Si(Li) detectors fabricated in-house~\cite{KP2018} and together with Shimadzu Corp~\cite{Kozai,Field}. Here, we report on the selection and testing of a passivation method to ensure the long-term stability of Si(Li) detectors for the GAPS experiment. GAPS is a balloon-borne instrument optimized to detect low-energy ($<$0.25\,GeV/nucleon) cosmic ray antinuclei. Several well-motivated dark matter (DM) models allow annihilation or decay into Standard Model particles, that would produce a flux of low-energy antinuclei~\cite{Donato2000,Donato2005,Duperray2005,Aramaki2016}. At low energies, secondary production of antideuterons and antihelium from cosmic ray interactions with the interstellar medium is suppressed, making low-energy antideuterons and antihelium a ``smoking gun" signal of dark matter~\cite{Mori2002,Hailey2009}. Using data from three Antarctic long duration balloon (LDB) flights, GAPS will set leading limits on the antideuteron and antihelium flux (at low energies)~\cite{AramakiSensitivity} and extend the antiproton spectrum to low energies (E$<$0.25\,GeV)~\cite{Aramaki2014}.

\par
\begin{figure}[h]
\centering
\includegraphics[width=3in]{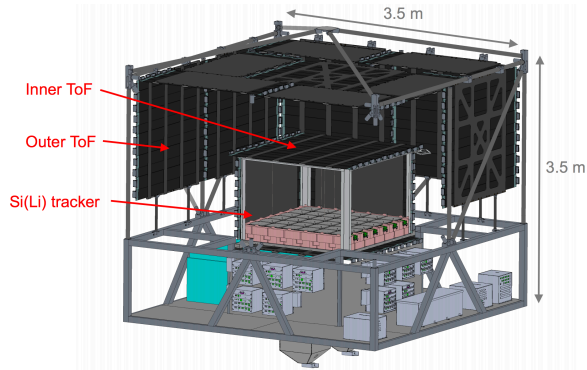}
   \caption{\label{f-payload}Exposed view of GAPS payload. Plastic scintillator paddles (black) comprise the outer TOF `umbrella' and inner TOF `cube.' The inner TOF cube surrounds the tracker, housing 10 layers of 144 detectors. Each detector module (gray) houses four Si(Li) detectors in 2x2 array. Structural supports are also shown.}
\end{figure}

 The GAPS instrument consists of a particle tracker and a time-of-flight (TOF) system (see Fig.~\ref{f-payload}). The TOF system is comprised of two layers of plastic scintillator that form an outer `umbrella' and an inner `cube' that are separated by $\sim$\,1\,m. The inner TOF cube encapsulates the Si(Li) particle tracker, providing $\sim$\,15\,m\textsuperscript{2} combined surface area of scintillator. The particle tracker consists of ten 1.6\,$\times$\,1.6\,m\textsuperscript{2} planes of Si(Li) detectors, with each layer separated with 10\,cm spacing. A large instrument is necessary to have high acceptance to cosmic ray antinuclei, but presents some experimental difficulties. Due to the large volume of the instrument, it is impossible to fly a pressure vessel or cryostat within the weight constraints of an LDB flight. Therefore, the Si(Li) detector array is cooled to relatively high operational temperatures, between $-$35 and $-$45\,\textdegree{C}, using a novel heat pipe system outlined in~\cite{Okazaki2018}. The Si(Li) detector array serves as both the target material to slow down the incoming antiparticle and the detector to measure an incoming particle's $\text d E/\text d x$ and the products of exotic atom de-excitation and annihilation.

\par The GAPS detection scheme exploits exotic atom physics to identify incoming antinuclei. A low-energy antiparticle first passes through the TOF, which measures its velocity and $\text d E/\text d x$ and provides a high-speed trigger. It then traverses the Si(Li) detector tracking system, undergoing d$E$/d$x$ losses until it is captured by an atomic nucleus in a Si(Li) detector or aluminum support, forming an exotic atom. The exotic atom de-excites, emitting Auger electrons and X-rays, and ultimately annihilates, producing pions and protons~\cite{AramakiXrayYields,Hartmann1990,Gotta2004}. The annihilation vertex provides a unique signature to discriminate antinuclei from baryonic cosmic rays. The stopping depth, $\text d E/\text d x$, energies of the X-rays produced during the de-excitation, and multiplicity of pions and protons emerging from the annihilation vertex are used to distinguish different antinuclei species.

\par This work reports on the R\&D effort to develop and select a passivation method for the GAPS Si(Li) detectors. Si(Li) passivation techniques and GAPS requirements for selecting a passivation method are reviewed in Sec.~\ref{s-review}. Adhesion and thermal testing of passivation candidates are described in Sec.~\ref{s-mechtesting}. In Sec.~\ref{s-noisetesting}, we outline the noise testing conducted to assess passivated detector performance. Accelerated lifetime testing and long-term monitoring of passivated detectors are reported in Sec.~\ref{s-protection}. Finally, conclusions and future prospects are presented in Sec.~\ref{s-conclusions}.

\section{Si(Li) Passivation Review \& Requirements\label{s-review}}

\par
\begin{figure}[h]
\centering
      \includegraphics[width=3in]{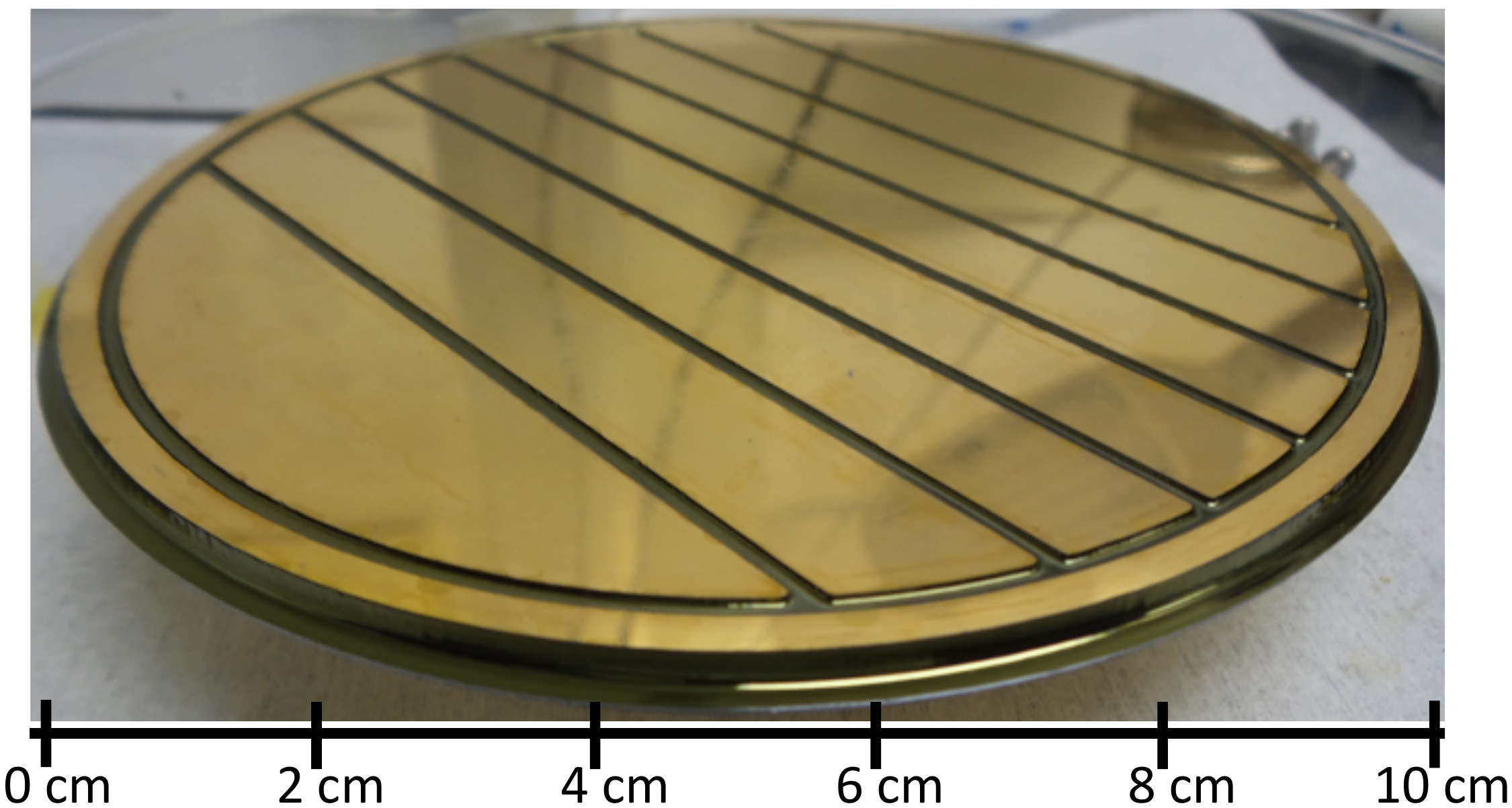}
      \includegraphics[width=3in]{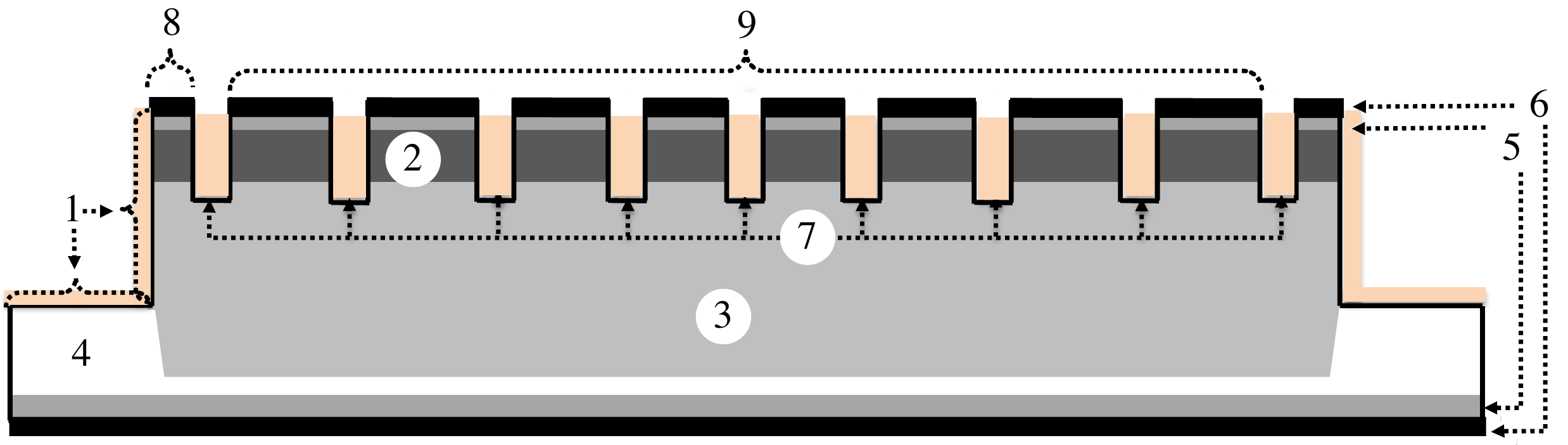}
      \caption{\label{f-detschematic}\textit{Top:} Photograph of passivated 8-strip GAPS flight detector with ruler for scale. If left unpassivated, the grooves segmenting the strips and guard ring are susceptible to contamination that can degrade detector performance. \textit{Bottom:} Diagram of 8-strip GAPS detector cross section. 1) The top hat geometry is defined using Ultrasonic Impact Grinding (UIG) to remove the top perimeter of the Si wafer, leaving a $\sim$\,1\,mm deep, $\sim$\,3\,mm wide top hat. 2) Li ions in the $\sim$\,100\,$\upmu$m diffused $n^+$ layer. 3) Li ions drifted into p-type bulk to create compensated active volume. 4) The Si in the top hat `brim' remains uncompensated. Electrodes consist of $\sim$\,20\,nm of Ni (5) and $\sim$\,100\,nm of Au (6). $\sim$\,1\,mm wide and $\sim$\,0.3\,mm deep grooves (7) segment the strips and separate the guard ring (8) from the active area (9). Orange shading indicates bare Si surfaces where a passivation coating is required.}
\end{figure}

 The GAPS Si(Li) fabrication method was developed in collaboration with Shimadzu Corporation, and further details on the fabrication technique and process yield are presented in~\cite{Kozai}. The geometry of the GAPS Si(Li) detectors is shown in Figure~\ref{f-detschematic}. In order to distinguish de-excitation X-rays from different antinuclei, the GAPS Si(Li) detectors must deliver $\lesssim$\,4\,keV FWHM energy resolution for 20$-$100\,keV X-rays. The LDB power and thermal constraints require the detectors to be operable at a bias voltage of 250\,V, at the relatively high temperatures of $-$35 to $-45$\,\textdegree{C}. The unpassivated detectors delivered the requisite noise performance, as demonstrated in~\cite{Field}. However, these measurements were made in a humidity-controlled lab environment and the detectors were cleaned immediately before testing. Detector performance is sensitive to the surface state of the exposed silicon of the grooves (Fig.~\ref{f-detschematic}, \#7) and top hat (Fig.~\ref{f-detschematic}, \#1)~\cite{goulding}. Specifically, the detectors are vulnerable to humidity and organic contaminants adsorbing onto the bare Si surface, and particulate matter (metal flakes from electrode, dust) that can fall into the grooves that segment the strips and guard ring. In order to achieve the GAPS goals, the detector performance must be stable over multiple years, so it was necessary to find a passivation method to protect the bare Si surfaces. 

\par GAPS requires a passivation candidate that provides a barrier to environmental contaminants and is robust to thermal cycling and mechanical shock. The passivation coating must protect the detector from its ambient environment and prevent the deleterious effects of surface contamination which can produce high leakage current and 1/f noise. The passivation process must be performed at low enough temperatures to avoid Li diffusion effects. Specifically, the Li in the $n^+$ layer (Fig.~\ref{f-detschematic}, \#2) must not extend below the detector grooves, which would effectively shunt the strips. Furthermore, the cure cycle must not cause significant decompensation of the positive Li ions in the p-type bulk, which could cause poor charge collection from the detector's active volume and increase the voltage required to deplete and operate the detector~\cite{goulding}. The passivation procedure must be adaptable to the geometrical constraints of GAPS, and be routinely applied by technicians to passivate the $>$\,1000 detectors needed for the GAPS flight.

\par The most commonly used passivation material for silicon radiation detectors is thermally grown silicon dioxide. However, optimal $\mathrm{SiO_2}$ layers are typically obtained at high temperatures ($T>1000$\,\textdegree{C}) in atmospheres of dry oxygen, wet oxygen or steam. Due to the high temperatures required, $\mathrm{SiO_2}$ was not explored. SiN, TaN, and TiN are often used as passivation coatings for Si substrates, and are typically produced using chemical vapor deposition (CVD), atomic layer deposition (ALD), or sputtering. These materials were not explored because the deposition techniques are not suitable for this work. CVD and ALD of SiN, TaN, and TiN typically require high substrate temperatures ($>$\,300\,\textdegree{C}), and sputtering is difficult to confine to the bare Si surfaces with high reproducibility~\cite{TiNReview,TaNReview,SiNReview_Kaloyeros_2017}. Another common method for passivation of silicon detectors is silicon monoxide. However, high temperature treatments are often necessary to produce optimal SiO films. Furthermore, previous studies indicate that devices passivated with SiO have higher leakage currents and additional 1/f noise~\cite{Hansen1980}. Thus SiO was not explored. Previous work demonstrated successful passivation of Si(Li) detectors using hydrogenated amorphous silicon ($\upalpha$-Si:H)~\cite{Walton}. $\upalpha$-Si:H is typically deposited by RF sputtering onto the detector. Amorphous silicon passivation was not explored because 1) the deposition process is complicated and reproducibility is poor, 2) the heat involved in the RF sputtering process can have an effect on Li compensation, and 3) GAPS operates Si(Li) detectors at higher temperatures than this previous study, and in this temperature range $\upalpha$-Si coated detectors have reported higher leakage current characteristics than bare detectors~\cite{Walton}.

\par Based on previous work, we chose to focus passivation R\&D on polymers, specifically polyimides and Parylene-C~\cite{Jantunen}. Previous work reports successful passivation of Si(Li) detectors using polyimides~\cite{Jantunen,Norm} and Parylene-C~\cite{Jantunen}. However, these detectors were operated at much lower temperatures than the GAPS operating temperature, where different noise components dominate. This work evaluates the viability of polyimide (PI) and parylene-C for passivation of Si(Li) detectors operated well above cryogenic temperatures.

\section{Mechanical Testing\label{s-mechtesting}}

\subsection{Motivation\label{ss-mechmotivation}}
In order for a passivation method to be viable for the GAPS experiment, the passivation coating must adhere to the Si surfaces. Polymer coatings typically fail due to cracking and/or delamination, and these failures are directly related to the state of stress in the coating. The stress is due to a combination of the material properties of the coating, the processing conditions used to produce the coating, and the environment. If the stress exceeds the ultimate strength of the coating, the coating fails by cracking. If the stored energy in the coating exceeds the work of adhesion to the substrate, the coating can delaminate. Furthermore, a mismatch in the coefficient of thermal expansion (CTE) between the polymer and the substrate can lead to delamination/cracking during temperature cycling~\cite{Sheth}. 

In order to test the mechanical properties of the coatings, adhesion and thermal testing was performed. Adhesion performance was assessed using a 180\textdegree{} pull test, following the ASTM D3359 standard (see Sec.~\ref{ss-at-success}). Thermal testing consisted of cycling the detector between room temperature and $-$50\,\textdegree{C}, and visually inspecting the polymer coatings under a microscope (see Sec.~\ref{sss-thermalcyclingmethod}).

\subsection{Sample Preparation\label{ss-sampleprep}}

\subsubsection{Wafer Cleaning and Preparation\label{sss-cleaning}}
Test-grade Si wafers\footnote{Test-grade Si wafers were procured from Addison Engineering and Wafer World}, were prepared for adhesion testing and thermal cycling studies. For adhesion testing studies, adherence to the ASTM D3359 standard required applying the polymer to a planar wafer surface. For thermal cycling studies, it was desired to apply the polymer to a sample geometry that is analogous to a detector, specifically to grooves that have been chemically polished to smoothness. Therefore, grooves were cut into thermal cycling samples using Ultrasonic Impact Grinding (UIG) before cleaning, etching, and applying the polymer (see Table~\ref{t-sampleprep}).

\par \textit{Thermal Sample UIG protocol:} Using UIG, 1 mm wide, 350 $\upmu$m deep grooves were cut into wafers. The groove depth and width were chosen to be analogous to the grooves segmenting the GAPS Si(Li) detectors. After UIG, the samples were cleaned ultrasonically in ACS-grade hexane to remove any wax and abrasive slurry from the UIG process. After ultrasonic cleaning with hexane, the following sample preparation was performed. 
\par \textit{Thermal \& Adhesion Test Sample Preparation:} A 3-step cleaning process was performed on all wafers, consisting of ultrasonic cleanings in ACS-grade acetone, methanol and DI water. All samples were etched in an HNA (20\,mL 49\% Hydrofluoric Acid, 35\,mL 60\% Nitric Acid, 55\,mL Glacial Acetic Acid) solution for 10~minutes. The etching process chemically polishes the Si surfaces, in particular the groove surfaces, which are left rough from the UIG process. With the etchant formulation used, a 10 minute etch was sufficient to produce the smooth and glassy groove surfaces typical of the GAPS Si(Li) detectors~\cite{Kozai}. These smooth surfaces presented an adhesion challenge, as rougher surfaces are typically better for polymer adhesion. After cleaning and etching, Polyimide or parylene-C was applied to the wafers. Adhesion and thermal testing was performed for two polyimides and Parylene-C, with and without an adhesion promoter.

\begin{table}[h]
\centering
 \caption{\label{t-sampleprep}Sample preparation and cleaning protocols for adhesion and thermal testing samples. 3-step cleaning entails ultrasonic cleaning a wafer in acetone, methanol, and DI for 5~minutes each, followed by drying with N2. Passivation coatings were applied to samples after etching.}
\begin{tabular}{|c|c|c|c|c|} 
 \hline
Sample Code & UIG & Hexane & 3-Step Clean & Etch \\ 
  \hline
 Adhesion & \xmark & \xmark & \cmark & \cmark  \\ 
  \hline
 Thermal & \cmark & \cmark & \cmark & \cmark  \\ 	
 \hline
\end{tabular}
\end{table}
\subsubsection{Polyimide Application\label{sss-piapp}}
\par Two polyimide (PI) precursors, VTEC PI-1388 and Ube U-Varnish-S, were used for the following studies. VTEC PI-1388 was selected due to its relatively low cure temperature ($\sim$\,250\,\textdegree{C} for full imidization), while Ube U-Varnish-S was selected because its CTE is a close match to silicon ($\sim$\,3\,ppm/\textdegree{C}). PI was applied by painting a polyimide precursor onto the wafer surface, and curing the polyimide in a Vulcan 3-1750 furnace. Each polyimide precursor is manufactured by combining a diamine and a dianhydride in a high polarity carrier solvent, typically N-Methyl-2-Pyrrolidine (NMP). The curing process drives the carrier solvent (NMP) out of the PI precursor, and facilitates the imidization (cyclization) of the diamine and dianhydride to form a PI. The final coating quality is extremely sensitive to the cure conditions, specifically the cure temperature and the heating rate. High cure temperatures (T\,$\geq250$\,\textdegree{C}) are desired to fully drive out the solvent and imidize the PI, but are not feasible for Si(Li) substrates because of high lithium mobility in silicon. To avoid movement of Li in the $n^+$ layer and compensated region, we did not use cure temperatures exceeding 210\,\textdegree{C}.

The following processing parameters were varied between PI samples: cure temperature, cure time, heating rate, use of silane adhesion promoter, and dilution of polyimide (see Table~\ref{t-piprotocols}). These processing parameters were tuned to produce PI coatings robust to mechanical and thermal stresses. Previous Si(Li) passivation literature suggested `soft-baking' the polyimide at 120\,\textdegree{C} for 25~minutes~\cite{Jantunen,Norm}, which drives most of the solvent from the PI precursor, but leaves the degree of imidization relatively low~\cite{Jantunen}. This soft-bake was used as a baseline cure condition, but further optimization was necessary based on adhesion and thermal testing results. `Rapid curing', by placing a substrate with PI into a pre-heated oven set to the cure temperature, was compared to `slow curing', by placing the substrate with PI into an oven at room temperature, and ramping the temperature to the cure temperature with a specified heating rate. For slow curing, a heating rate of 5\,\textdegree{C}/min was used. This heating rate was chosen based on a previous study that found that heating rates $<$\,10\,\textdegree{C}/min lead to a higher degree of solvent removal and imidization~\cite{SolventRemoval}. To test the effect of silane on polyimide adhesion, $\gamma$-Aminopropyltriethoxysilane (APS), was applied to some etched wafers before applying the PI. A 0.1\% (v/v) solution of APS was prepared in DI water and mixed for 1~hour on a hot plate with a magnetic stir bar. The APS solution was then applied by hand to the clean wafer surfaces, and baked on a hot plate at 85\,\textdegree{C} for 30~minutes. This APS application protocol was based on a previous study, which demonstrated that APS increased the adhesion strength of the PI-silicon interface by a factor of 25~\cite{Sheth}. The PI precursor was either applied to the substrate `neat' as it arrived from the manufacturer, or in a 1:1 dilution of PI precursor and pure NMP. Diluting the PI precursor solution decreases its viscosity, and enables application with a pipette.

\begin{table}[h]
\centering
\caption{\label{t-piprotocols}PI application parameters were optimized to provide a coating with good adhesion and thermal properties. Initial samples (Sample code 1A) were `rapidly cured' by placing them in an oven pre-heated to the cure temperature; however, it was quickly determined that `slow curing' by ramping the oven temperature from room temperature to the cure temperature yielded better coatings. The final PI passivation protocol is highlighted in gray.}
\begin{tabular}{|c|c|c|c|c|c|} 
 \hline
Sample Code &Cure Temperature& Cure Time & Heating Rate& APS & Dilute PI  \\ 
  \hline
 1A & 120 \textdegree{C} & 25\,mins & None & \xmark & \xmark \\ 
    \hline
 1B & 120 \textdegree{C} & 25\,mins & 5\,\textdegree{C}/min & \xmark & \xmark \\ 
  \hline
 2A & 180 \textdegree{C} & 10\,mins & 5\,\textdegree{C}/min & \xmark & \xmark \\ 
 \hline
 2B & 180 \textdegree{C} & 10\,mins & 5\,\textdegree{C}/min & \cmark & \xmark\\ 
 \hline
  2C & 180 \textdegree{C} & 10\,mins & 5\,\textdegree{C}/min & \xmark & \cmark\\ 
 \hline
 2D & 180 \textdegree{C} & 10\,mins & 5\,\textdegree{C}/min & \cmark & \cmark\\ 
 \hline
  3A & 180 \textdegree{C} & 25\,mins& 5\,\textdegree{C}/min & \xmark & \xmark \\ 
 \hline
 3B & 180 \textdegree{C} & 25\,mins & 5\,\textdegree{C}/min & \cmark & \xmark \\ 
  \hline
  3C & 180 \textdegree{C} & 25\,mins& 5\,\textdegree{C}/min & \xmark & \cmark \\ 
 \hline
 3D & 180 \textdegree{C} & 25\,mins & 5\,\textdegree{C}/min & \cmark & \cmark \\ 
 \hline
  4A & 210 \textdegree{C} & 60\,mins& 5\,\textdegree{C}/min & \xmark & \cmark \\ 
 \hline
 \cellcolor{gray!60}4B & \cellcolor{gray!60}210 \textdegree{C} & \cellcolor{gray!60}60 mins & \cellcolor{gray!60}5 \textdegree{C}/min &\cellcolor{gray!60} \cmark &\cellcolor{gray!60} \cmark \\ 
 \hline
\end{tabular}
\end{table}

\subsubsection{Parylene-C Application}
Parylene-C was applied in a vapor deposition chamber (SCS Labcoter 2, PDS 2010). The deposition process conformally coats the substrate with a Parylene-C film. For a given deposition chamber, the resulting film thickness is proportional to the mass of the dimer that is vaporized, so that for the chamber employed in this study 1\,g of dimer resulted in a 1\,$\upmu$m thick coating. For all parylene-C samples, 5\,g of dimer was used to produce 5\,$\upmu$m thick coatings. For some samples, a silane adhesion promoter (A-174) was applied before parylene-C deposition, to assess the silane's impact on coating adhesion. For adhesion and thermal testing studies, samples were conformally coated.

In preparation for noise testing, a method to mask the readout strips was developed. Selective deposition by shadow masking and surface priming was attempted. For shadow masking, wafers were wedged in between two aluminum plates with machined cut-outs in the shape of the detector strips. After parylene-C deposition, the masks were cut away from the wafer using a sharp blade. For surface priming, a Micro-90 solution was painted onto the strip surface before parylene-C deposition. Micro-90 (Mfg: Cole-Parmer) is a soap-like solution that inhibits the Parylene-C adhesion to silicon, enabling us to peel the Parylene-C from the selectively primed surfaces~\cite{Micro90}. After vapor deposition, the Parylene-C was mechanically removed from the electrodes by scraping and pulling with electrostatic discharge safe polyvinylidene fluoride tipped tweezers and the Micro-90 was cleaned from the detector surface using a methanol-soaked swab.

\subsection{Adhesion Testing \& Results}
\subsubsection{\label{ss-at-success}Adhesion Testing Method \& Success Criteria}
Adhesion testing was performed on test samples using an ASTM D3359 cross hatch adhesion test~\cite{D3359}. Using a razor blade, an X-shape cut was notched into the polymer coating. Elcometer 99 tape was pressed and smoothed down onto the coating surface, on top of the X-cut. Within 90\,$\pm$\,30 seconds of applying the tape, the tape was pulled off, pulling it back upon itself at a 180\textdegree{} angle. After pulling, the coating was inspected for blemishes, and the tape was inspected for residue from the coating. The degree of polymer removal due to the pull test is graded on a 0-5 scale (5 indicates no polymer removed, 0 indicates a majority of the polymer was removed). In order to pass adhesion testing, a coating had to pass the 180\textdegree{} pull test with a score greater than four.
\subsubsection{Adhesion Testing Reults\label{sss-atresults}}
\par No difference in adhesion strength was noted between neat and dilute samples of a given PI. PI samples without APS failed all adhesion tests, while PI samples with an APS pre-coating passed all adhesion tests, for both neat and dilute PI-1388 and U-Varnish-S. Without an APS adhesion layer, VTEC PI-1388 demonstrated stronger adhesion properties than Ube U-Varnish-S, as U-Varnish-S samples demonstrated a higher degree of PI removal after adhesion testing, regardless of dilution. All samples prepared with the final PI application protocol (Sample Code 4B) passed adhesion tests with a grade of 5.

\par Conformally coated parylene-C samples passed all adhesion tests, with and without a silane base layer. Thus, for noise testing, parylene-C samples were not prepared with a silane pre-coating, while PI samples were primed with an APS / DI solution.

\subsection{Thermal Testing \& Results}
\subsubsection{Thermal Cycling Testing Method \& Success Criteria\label{sss-thermalcyclingmethod}}
Samples were thermal cycled in a custom testing setup that consisted of dry ice and EPS foam insulation. The testing setup was assembled to cycle the samples between room temperature and $-$50\,\textdegree{C}, with a ramp rate $<$\,5\,\textdegree{C}/min. This temperature profile is consistent with the cooling used during laboratory calibration and expected during the LDB flight. During thermal cycling, the wafers were kept in polypropylene carrying cases, to avoid condensation on the wafer surfaces when opening the chamber. After each thermal cycle, the coatings were inspected visually and under a microscope for cracking and delamination. A sample was required to survive twelve thermal cyclings without exhibiting delamination or cracking to be deemed successful.
\subsubsection{Thermal Cycling Results}
Parylene-C coatings were extremely robust to thermal stresses, and did not demonstrate any delamination or cracking through 12 thermal cycles. 

\begin{figure}[h]
\centering
\hspace*{\fill}%
\centering
\begin{minipage}{.35\textwidth}
\centering
\vspace{0pt}
\includegraphics[width=\linewidth]{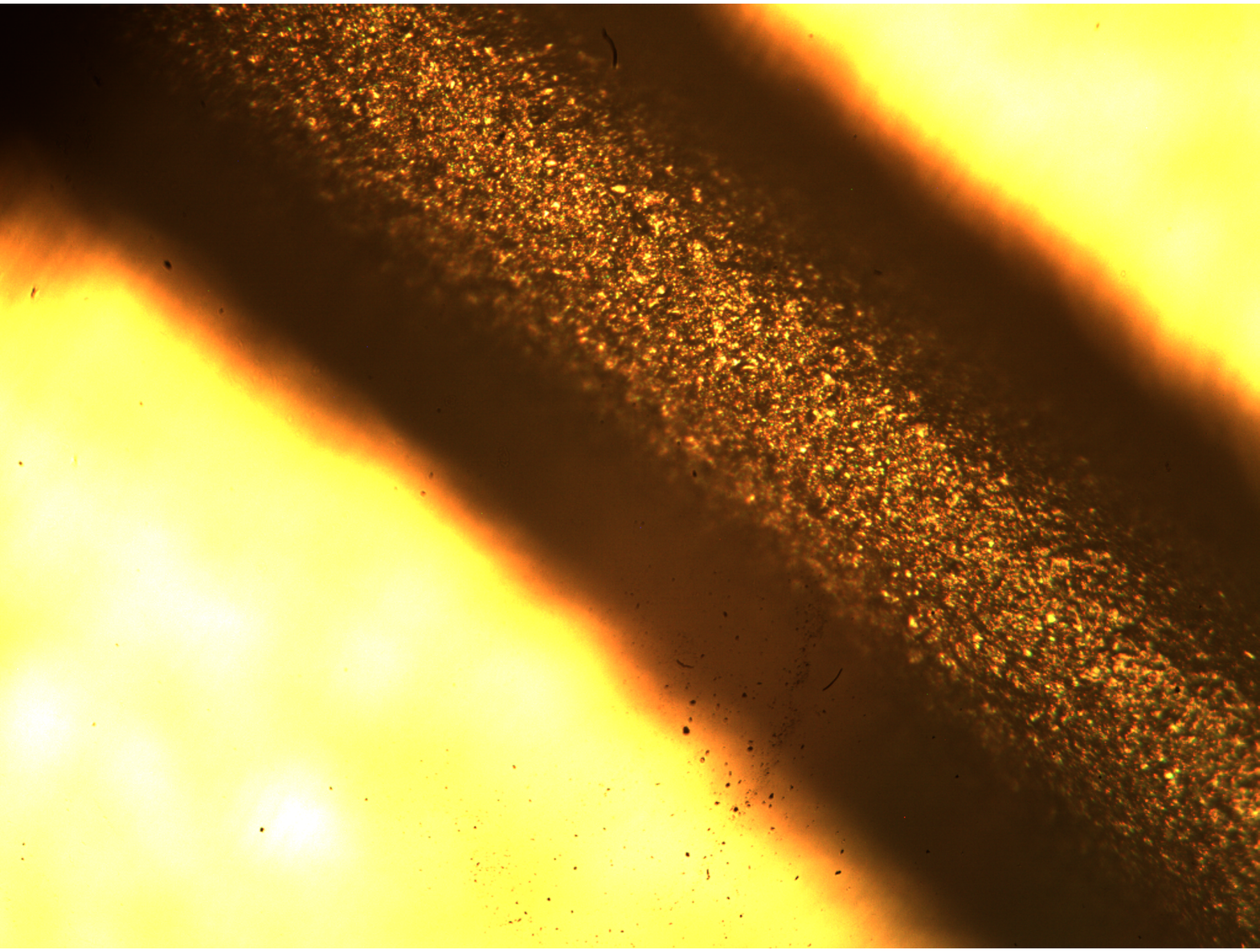}
\end{minipage}%
\begin{minipage}{.35\textwidth}
\centering
\includegraphics[width=\linewidth]{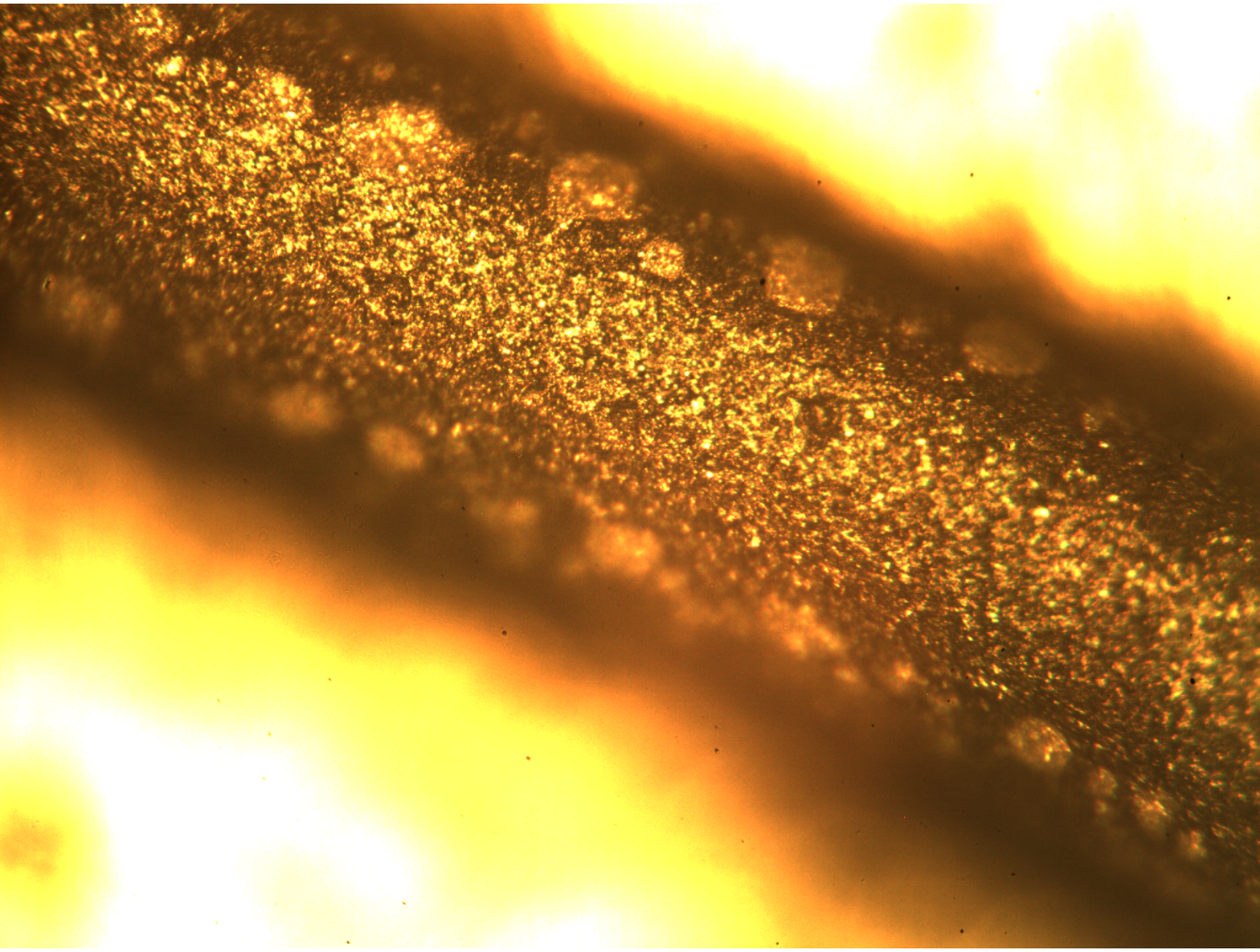}
\end{minipage}
\begin{minipage}{.2\textwidth}
\centering
\includegraphics[width=\linewidth]{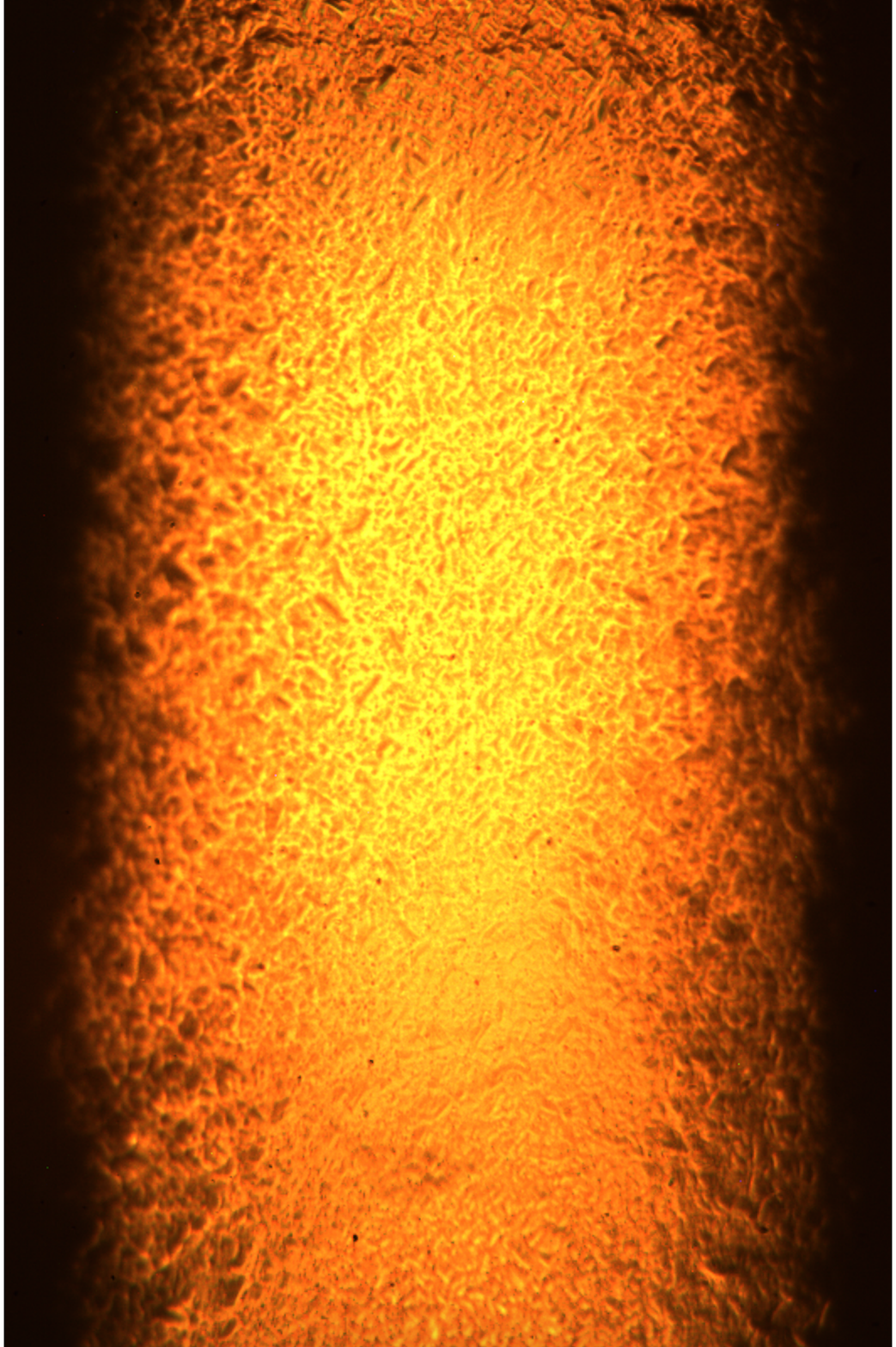}
\end{minipage}
\captionof{figure}{\textit{Left/Center:}\label{f-thermalcycling} Microscope images of grooves of thermal test sample before (left) and after (center) two thermal cycles. After two thermal cycles, spots appeared at bottom of groove. These spots are indicative of delamination of the PI from the silicon surface, due to CTE mismatch. This sample was cured at 120\,\textdegree{C} for 25~minutes, a cure cycle that was found to be insufficient to produce a robust coating. \textit{Right:} Microscopic images of a PI coated groove, passivated using the final procedure. No delamination or cracking was observed after twelve thermal cyclings in PI coatings using the final passivation procedure.}
\end{figure}

Initially, PI samples were cured at 120\,\textdegree{C} for 25~minutes (Sample Codes 1A-1B), based on previous successful passivation work~\cite{Jantunen,Norm}. However, these samples exhibited delamination and cracking through successive thermal cyclings, and the cure temperature was subsequently increased to 180\,\textdegree{C} (Code 2A-3D). PI applied `neat' to the grooves demonstrated poor reproducibility, as the neat PI precursor had a high contact angle with the Si surface and aggregated during curing, leaving bare silicon surfaces exposed. Therefore, dilute PI application was selected for noise testing. Rapidly cured samples, cured in an oven pre-heated to the cure temperature, demonstrated higher failure rates than slow cured samples because slow curing promotes better solvent removal and prevents thermal shock in the coating~\cite{SolventRemoval}. The PI samples primed with APS and slow cured at $\geq$180\,\textdegree{C} for $\geq$10 mins (Sample Codes 2D, 3D, 4B) were robust to thermal cycling (see Fig.~\ref{f-thermalcycling}).

\section{Noise Testing\label{s-noisetesting}}
\subsection{Motivation}
After tuning the application protocols to produce coatings with acceptable adhesion and thermal properties (see Sec.~\ref{s-mechtesting}), the passivation coatings were applied to Si(Li) detectors, and detector noise performance was assessed before and after passivation. The main requirement is that GAPS Si(Li) detectors must provide $\lesssim$\,4\, keV energy resolution to 20--100\,keV X-rays. Calibration measurements are made using a discrete preamplifier, whereas during final calibration and the LDB flight a custom ASIC will be used~\cite{ASIC}. Therefore, it is essential to understand the noise characteristics of each detector (leakage current, capacitance, resistance, etc.) to simulate detector energy resolution with the ASIC design parameters for the LDB flight. A noise model was used, and the noise model fit parameters were included in the success criteria, which are outlined in Sec.~\ref{ss-noisetestingsetup}.

\subsection{Detector Preparation\label{ss-detprep}}
A detector's grooves and top hat were cleaned before passivation. The standard cleaning protocol consists of applying methanol to the tip of a cleanroom swab, and gently swabbing the detector's bare silicon surfaces. This cleaning protocol removes any particulate contamination from the surfaces and sets the surface state of the silicon to be lightly n-type. The cleaning must be performed in a low-humidity environment ($<$10\% relative humidity), specifically a N2 purged glove box, to yield consistently good results. Detectors fabricated in-house at Columbia University (TDxxxx) and at Shimadzu (Shxxxx) were passivated and tested. There are some differences in the fabrication protocols~\cite{KP2018,Kozai}. Typically, the in-house detectors were used to demonstrate proof of concept, whereas the success criteria were ultimately applied to Shimadzu detectors.

Parylene-C and PI passivation coatings were applied to in-house detectors as proof of concept. Parylene-C was applied using Micro-90 to mask the electrodes and enable readout (see Sec.~\ref{sss-passoptimization}). PI was applied using an APS adhesion promoter and a diluted VTEC PI-1388 polyimide precursor solution. For initial testing purposes, the APS and PI cure conditions were varied to produce optimal passivated detector noise performance. The minimal cure temperature and time that produced coatings with acceptable adhesion and thermal properties were 85\,\textdegree{C} for 30~minutes for APS curing, and 180\,\textdegree{C} for 10~minutes for PI curing (see Sec.~\ref{sss-piapp}). These cure conditions were compared to higher temperature cures that were chosen to exceed the boiling point of the solvent in the APS and PI precursor solution. To improve the solvent removal and degree of PI imidization, PI was cured at 210\,\textdegree{C} for 1~hour, and APS was cured at 110\,\textdegree{C} for 20~minutes. Based on the initial testing results, where lower temperature cures resulted in degraded leakage current and energy resolution (see Sec.~\ref{sss-passoptimization}), the higher temperature cures were used for all subsequent PI passivated detectors.

\par The final PI passivation protocol is as follows:

{\centering
\fbox{\begin{minipage}[h]{\linewidth}
\centering
\begin{itemize}[noitemsep,topsep=2pt]
\item Mix 0.1\% (v/v) solution of APS in de-ionized water for 1~hour
\item Apply APS solution to detector grooves and top hat using pipette
\item Bake detector in open glass petri dish on hotplate at 110\,\textdegree{C} for 20~minutes
\item Let detector cool, mix 1:1 dilution of PI precursor in NMP by hand with teflon applicator
\item Degas PI in rough vacuum for 10~minutes to remove bubbles
\item Apply PI precursor solution to detector grooves and top hat using pipette
\item Cure in oven at 210~\textdegree{C} set point temperature for 1~hour with 5\,\textdegree{C}/min heating rate
\item After 1~hour at 210\,\textdegree{C} set point, prop oven open to decrease temperature gradually
\item When oven temperature reaches 70\,\textdegree{C} ($\sim$\,40~minutes), remove detector from oven and place in dry box
\end{itemize}
\end{minipage}}
}
 
\par Temperature testing was conducted to ensure that the substrate temperature reached 110\,\textdegree{C} during the APS cure and $>$\,204\,\textdegree{C} during the PI cure. A dummy substrate was placed in the oven and on the hot plate used for curing passivation coatings. Thermocouples were used measure the substrate's surface temperature. A MicroDAQ USB-TEMP temperature data acquisition module was used to log the substrate temperature at each second during the cure cycles. During APS curing, the substrate equilibrated to $\sim$\,110\,\textdegree{C} after 10~minutes. In the oven, the substrate reached $\sim$\,210\,\textdegree{C} at the end of the 1~hour cure cycle, and the temperature ramp rate was kept below 5\,\textdegree{C}/min to avoid thermally shocking the PI coating.

\subsection{Noise Testing Set-up\label{ss-noisetestingsetup}}
Noise testing was performed at MIT. To assess the success of a passivation coating, room temperature leakage current and cold ($\sim$\,--37\,\textdegree{C}) spectral measurements were performed on a detector before and after applying the passivation coating. Per-strip leakage current was measured directly using a Keithley 487 Picoammeter / Voltage source with all other strips and guard ring grounded. The Keithley 487 ramped the voltage applied to the p-side from 0\,V to $-$400\,V in 25\,V increments while measuring the resulting leakage current from a given strip on the n\textsuperscript{+} side.

Energy resolution measurements were performed in a custom vacuum testing chamber outlined in~\cite{Field} or in a SUN EC13 Temperature Chamber. Both chambers were cooled using LN2; the custom chamber was pumped to $\sim$\,2\,Pa during operation, whereas the SUN chamber was purged with nitrogen. Detector temperature was monitored using a calibrated diode, and spectral measurements were recorded in the typical operating temperature range, $-$35\,\textdegree{C}\,$>$\,T\,$>-$45\,\textdegree{C}. In the chambers, the detectors were uniformly irradiated with a 100\,$\mathrm{\upmu Ci}$ \textsuperscript{241}Am radioactive source. During operation, the p-side of a detector was biased to $-$250\,V using a Tennelac 953 HV supply in the vacuum chamber, and a CAEN N1471 in the SUN chamber. The signal was readout from the n\textsuperscript{+} side by a custom 8-channel discrete-component charge-sensitive preamplifier board, which was pressure mounted to the strips via spring-loaded pins. In the vacuum chamber, signal from one preamplifier was processed by a Canberra 2020 Spectroscopy Amplifier at various peaking times and digitized by an Ortec Ametek Easy MCA module. In the SUN chamber, signals are shaped and digitized using a CAEN N6725 digitizer.

\par A noise model is used to characterize each detector, identify each noise source that contributes to the overall energy resolution, and determine if the noise arises from intrinsic detector performance or from the readout chain~\cite{Field,2012NIMPA.682...90A}. The equivalent noise charge (ENC) from the detector and readout chain, and the FWHM energy resolution can be estimated as follows~\cite{goulding,Spieler2005}:

\begin{equation}\label{eq:noisemodel1}
ENC^2 = \left( 2qI_{leak}+\frac{4kT}{R_{p}} \right)\tau F_{i} + 4kT \left(R_{s} + \frac{1}{g_{m}} \right) \frac{C_{tot}^2}{\tau} F_{\nu} + A_{f}C_{tot}^2 F_{\nu f}
\end{equation}

\begin{equation}\label{eq:noisemodel2}
FWHM=2.35 \epsilon \frac{ENC}{q}
\end{equation}

In Eqs. \eqref{eq:noisemodel1} and \eqref{eq:noisemodel2}, $q$ is the fundamental charge, $k$ is the Boltzmann constant, and $\epsilon$ is the ionization energy of silicon (3.6\,eV per electron-hole pair). $R_{p}$, $g_{m}$, $F_i$, $F_{\nu}$, and $F_{\nu f}$ are fixed parameters specific to preamplifier and shaping amplifier used in the readout chain and are outlined in~\cite{Field}. 

\par The most relevant parameters in this study are $I_{leak}$, $C_{tot}$, $A_{f}$, and $R_{s}$, which are fit to characterize the noise performance of the detector. $I_{leak}$ is the temperature-dependent per-strip leakage current. The total input capacitance ($C_{tot}$) is the sum of all parallel capacitances including the individual strip capacitance ($C_{det}$), the capacitance of the FET input stage ($C_{FET}$\,$\sim$\,10\,pF), the interelectrode capacitance of adjacent strips and the grounded guard ring ($C_{int}$), and any stray capacitance ($C_{stray}$\,$\sim$\,5\,pF). $R_{s}$ is the  sum of all series resistances that can arise from the detector and preamplifier mounting. $A_{f}$ is the coefficient of 1/f noise which may arise from surface effects or (ideally just) preamplifier noise.

\par Energy resolution was measured as a function of peaking time for detectors at a given operational temperature before and after applying a passivation coating. The peaking time ($\tau$) vs. energy resolution (FWHM) data is used with Eq.~\eqref{eq:noisemodel2} to find a best fit for $I_{leak}$, $C_{tot}$, $A_f$, and $R_{s}$ while keeping the other noise model parameters fixed. Since, $C_{tot}$, $A_f$, and $R_{s}$ are degenerate, they cannot be fit simultaneously, so an iterative approach is used, which is described in~\cite{Field}. The fitted values for each strip are compared before and after passivation, and we assess the success criteria based on the overall energy resolution and fit parameters.

\par The following success criteria were imposed on passivated detector energy resolution and noise model fit parameters: 

\begin{enumerate}
\itemsep 0em 
\item FWHM energy resolution at optimal peaking time $\lesssim$\,4\,keV
\item $I_{leak}\lesssim 4$\,nA
\item $A_{f}\lesssim 2.5 \times 10^{-13}$\,V\textsuperscript{2}. 
\end{enumerate}
The fitted $I_{leak}$ and $A_f$ cutoffs were chosen based on unpassivated detector measurements performed in~\cite{Field}.

\subsection{Noise Testing Results}
\subsubsection{Passivation Protocol Optimization\label{sss-passoptimization}}
Before passivating Shimadzu flight detectors, in-house fabricated detectors were passivated and tested to validate a passivation method. Parylene-C was applied to an in-house fabricated Si(Li) detector, TD0087. After applying the parylene-C coating using selective deposition with Micro-90, the detector's leakage current saturated the picoammeter's current limit (2.5\,mA) at $\sim$\,30\,V. Since the leakage current from the active area was $<$\,1\,$\upmu$A during this test, it indicated a large leakage current through the guard ring due the passivation coating. Based on this result and concerns about the reproducibility of the coating method, Parylene-C was not further explored.

An in-house fabricated Si(Li) detector, TD0093, was passivated with PI cured at 180\,\textdegree{C} for 10~minutes. After passivation, the detector's leakage current was an order of magnitude higher than its pre-passivation value at room temperature, $-$36\,\textdegree{C}, and $-$48\,\textdegree{C}. Re-baking the detector at 180\,\textdegree{C} for 25~minutes reduced the detector leakage current to its pre-passivation values.  Therefore, we determined that the elevated leakage current was caused by an insufficient cure cycle that did not drive out the solvent and imidize the PI precursor, and all subsequent detectors passivated with PI were cured at 210\,\textdegree{C} for 1~hour. The heating involved in this cure cycle did not significantly affect the detector's lithium distribution, and did not degrade detector energy resolution when operated at 250\,V. We note that the voltage required to deplete and operate the detector increases from $\sim$\,80\,V before passivation, to $\sim$\,150\,V after passivation; however, this is acceptable since the detectors will be operated at 250\,V.
\par
\begin{figure}[H]
\hspace*{\fill}%
\centering
\begin{minipage}{.5\textwidth}
\centering
  \includegraphics[width=\linewidth]{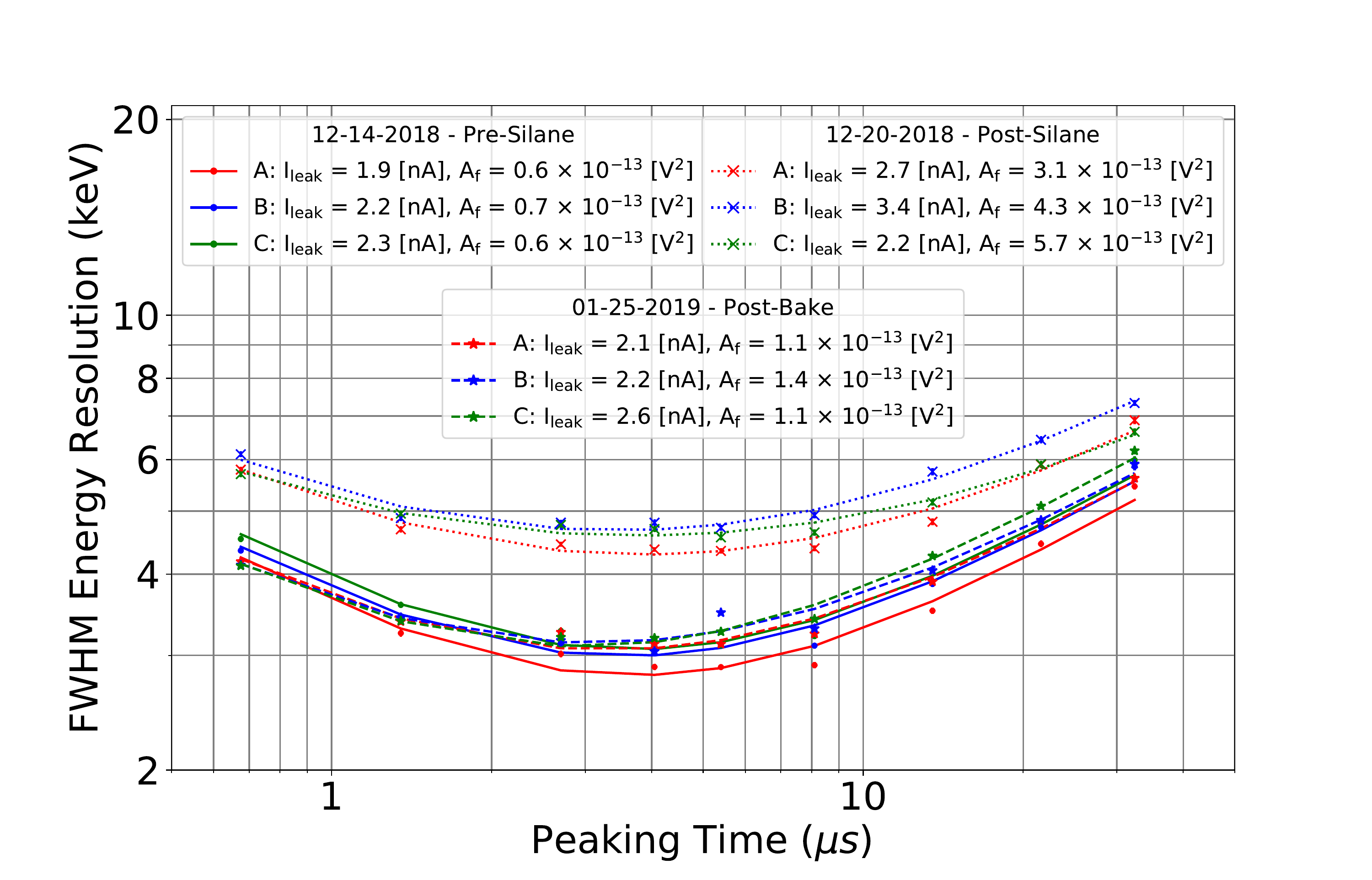}

\end{minipage}%
\begin{minipage}{.5\textwidth}
\centering
\includegraphics[width=\linewidth]{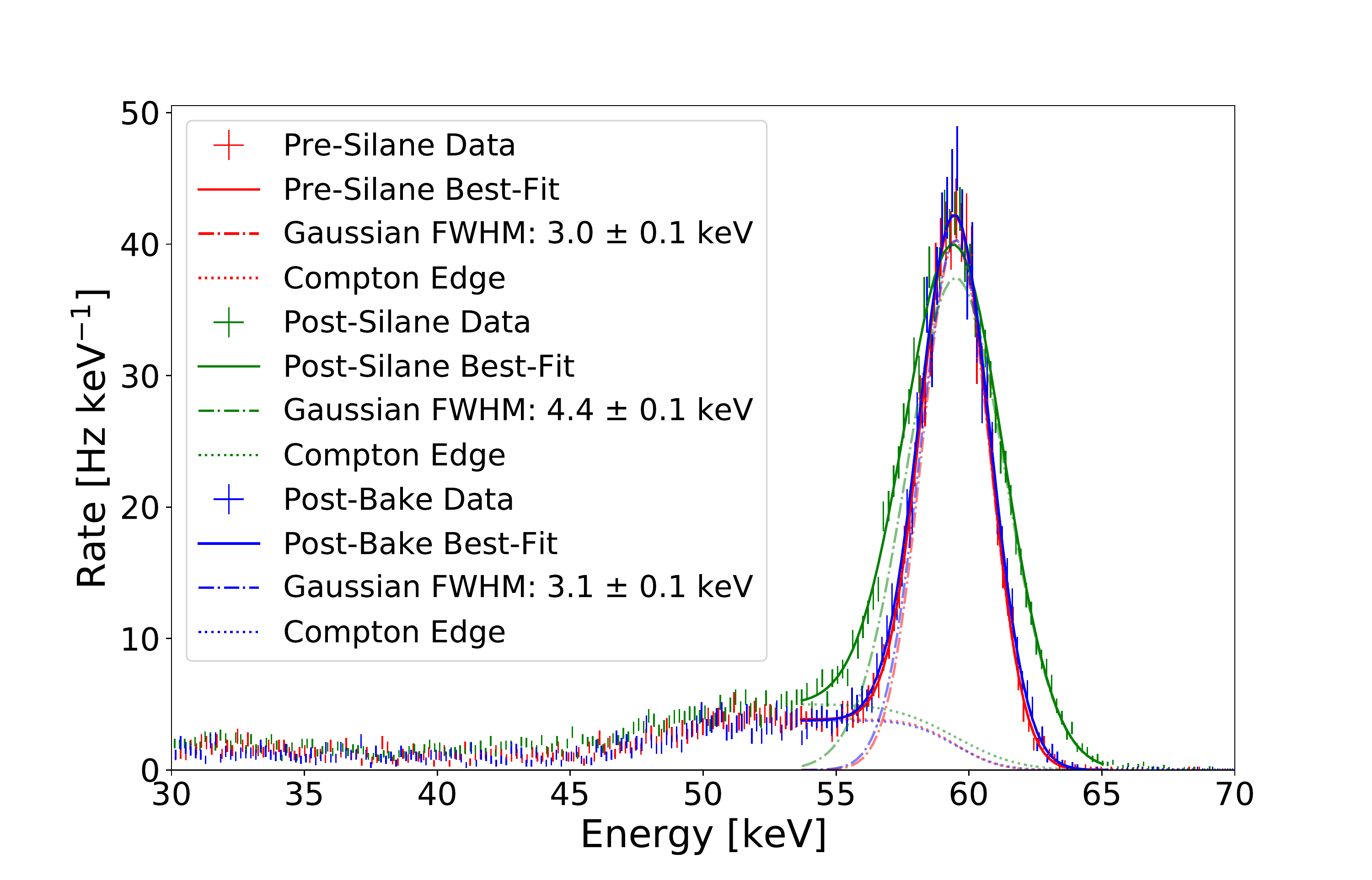}
\end{minipage}%

 \captionof{figure}{\label{f-silanebake}\textit{Left:}~Energy resolution (FWHM) as a function of peaking time for three strips of the 8-strip detector Sh0075, measured at $\sim$\,--37\,\textdegree{C}. Baseline measurements were taken after cleaning the detector strips as outlined in Sec.~\ref{ss-detprep}. Post-silane measurements were taken after applying the APS adhesion promoter to the detector's grooves and top hat and curing on a hot plate at 85\,\textdegree{C} for 30~minutes. After this APS cure cycle, the detector's energy resolution degraded and 1/f noise component increased. After re-baking the detector at 110\,\textdegree{C}, the energy resolution and 1/f noise component recovered to pre-passivation levels.~\textit{Right:}~Corresponding spectra for strip A of Sh0075 at the optimal peaking time (4\,$\upmu$s). Each spectra shows a photopeak and a low-energy tail. GEANT4 simulations have confirmed that the low-energy tail is due to X-rays scattered from material in the testing chamber and are not due to charge trapping. The data is fit to a Gaussian (dash-dotted) plus an error function (dotted) as discussed in~{\cite{Field}}. Pre-silane and post-bake measurements were performed using a preamplifier with higher attenuation than the preamplifier used for post-silane measurements, so the post-silane count rate was scaled to make the difference in spectral shape more evident.}
\end{figure}

 To isolate the effects of the APS application procedure, APS was applied to an eight-strip Shimadzu detector, Sh0075, following the protocol used for thermal and adhesion samples. After curing at 85~\textdegree{C} for 30~minutes, the detector's spectral performance degraded due to 1/f noise (see Fig.~\ref{f-silanebake}). Re-heating the detector at 110\,\textdegree{C} for 20~minutes improved the detector's spectral performance so that no degradation was noted compared to pre-passivation measurements. Therefore, the 85\,\textdegree{C} cure was deemed insufficient to dry the APS solution, and APS was cured at 110\,\textdegree{C} for all subsequent passivations.

\subsubsection{Energy Resolution Testing and Large-Scale Validation\label{sss-eresvalidation}}

\begin{figure}[H]
\hspace*{\fill}%
\centering
\begin{minipage}{.5\textwidth}
\centering
\vspace{0pt}
\includegraphics[width=\linewidth]{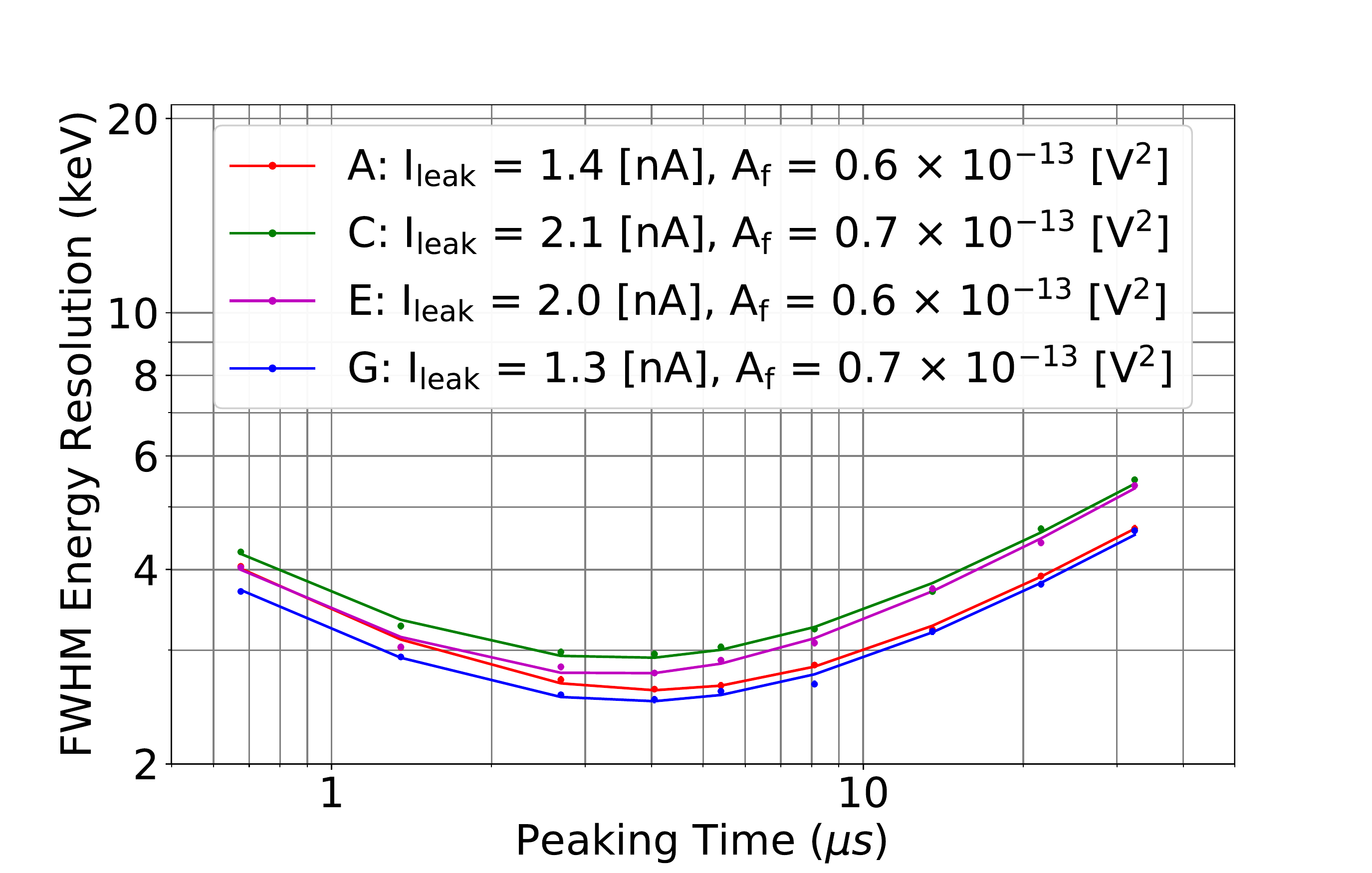}
\end{minipage}%
\begin{minipage}{.5\textwidth}
\centering
\includegraphics[width=\linewidth]{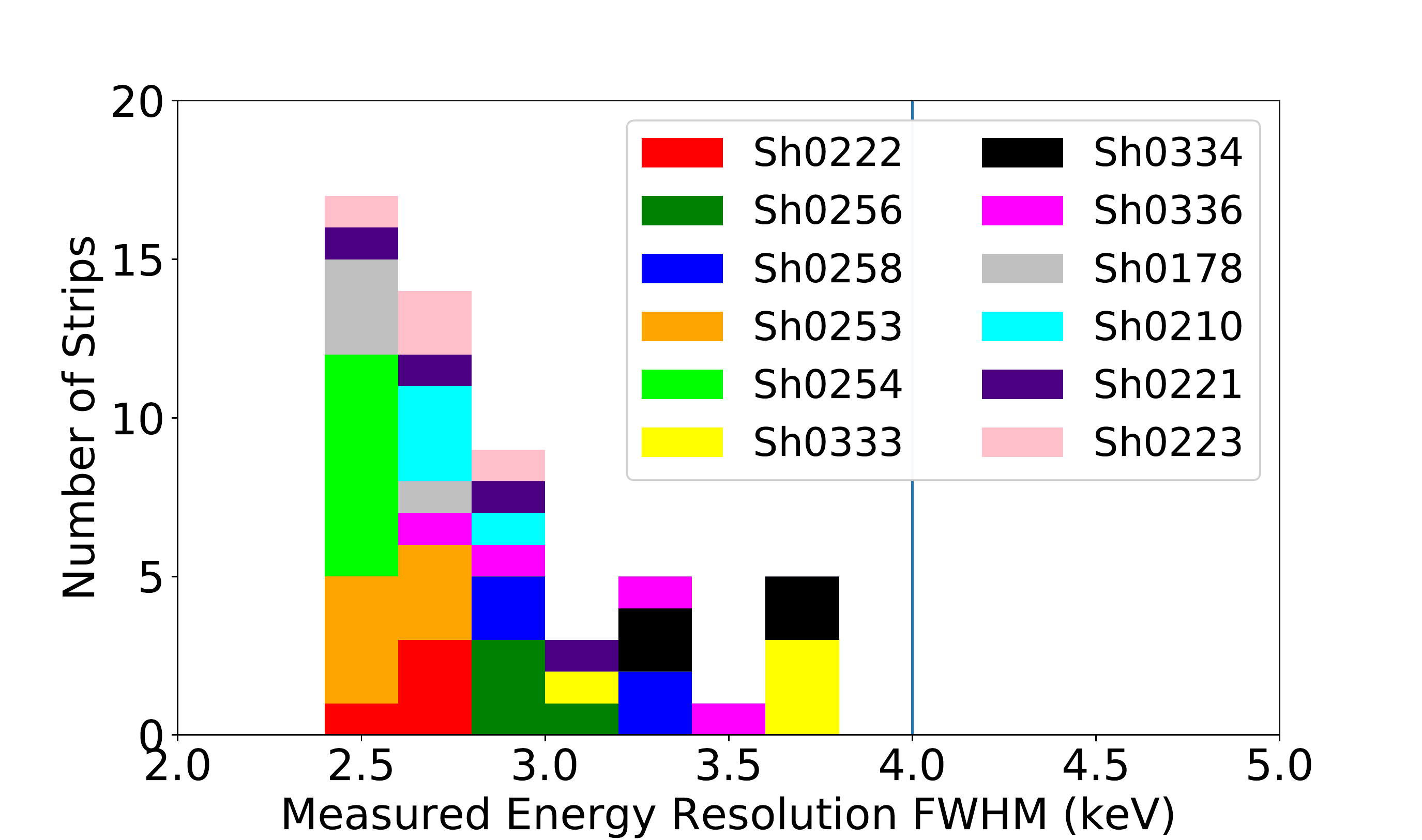}
\end{minipage}
\centering
\begin{minipage}{.5\textwidth}
\centering
\includegraphics[width=\linewidth]{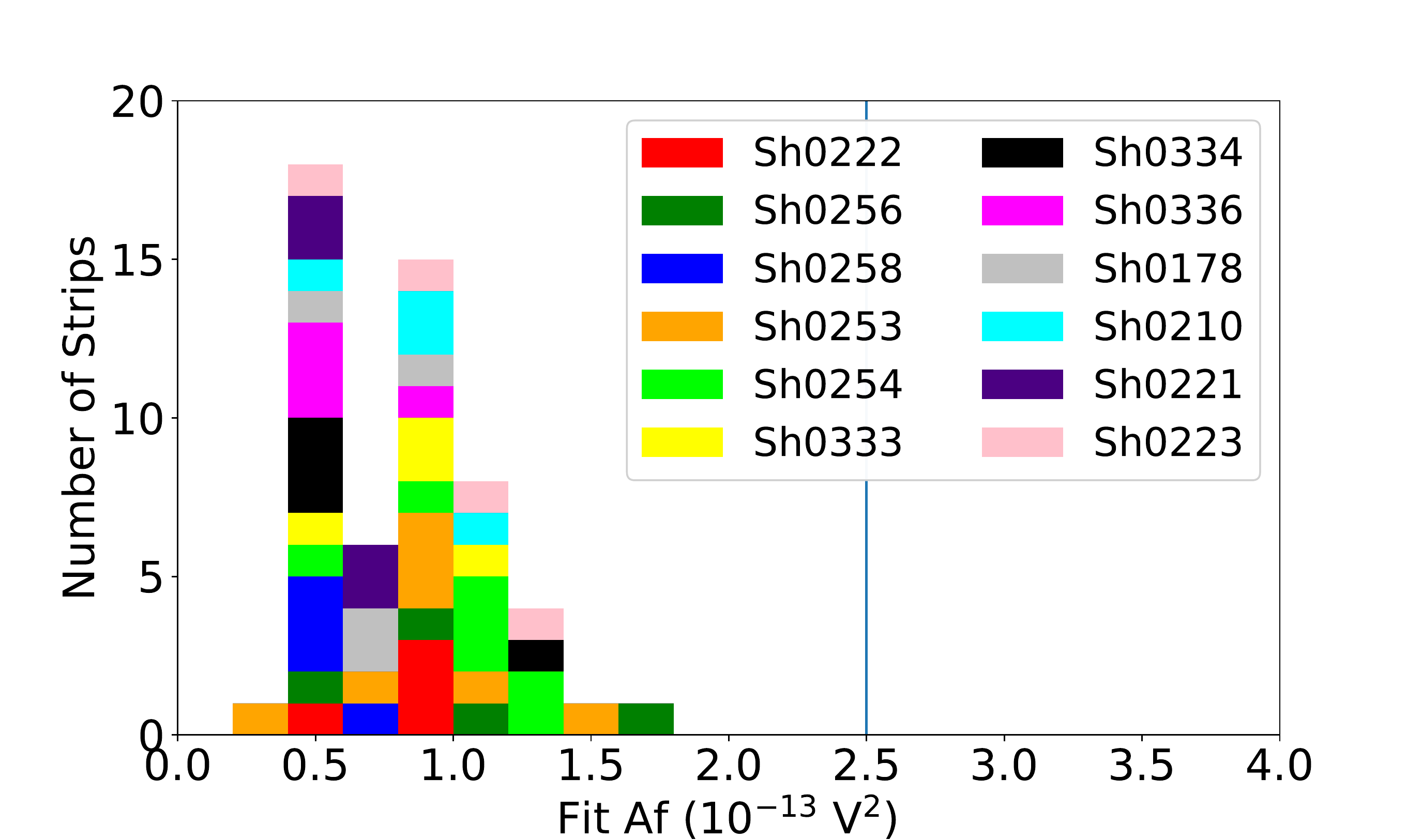}
\end{minipage}%
\begin{minipage}{.5\textwidth}
\centering
\includegraphics[width=\linewidth]{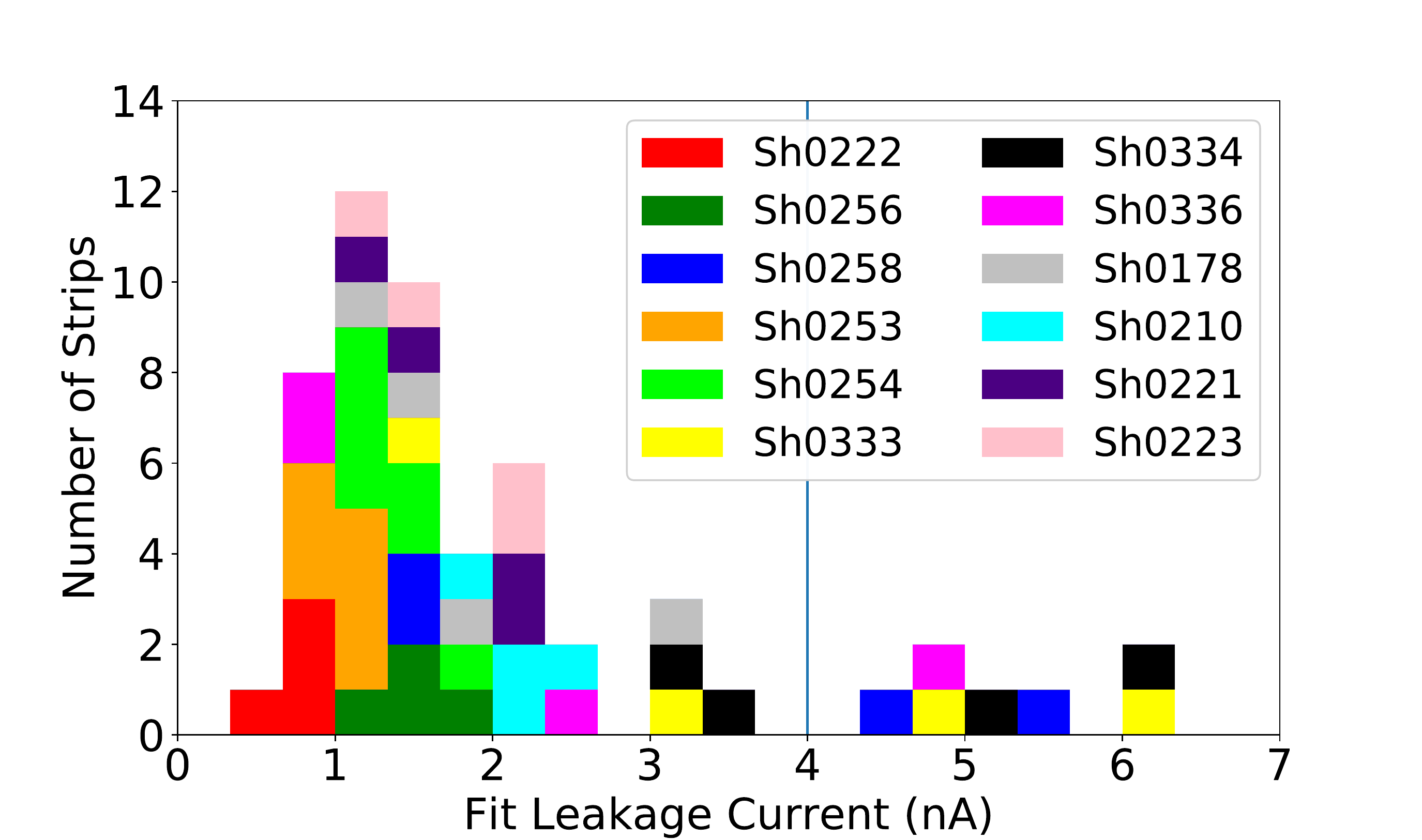}
\end{minipage}
\label{fig:test4}
 
\captionof{figure}{\label{f-passval}\textit{Top Left:} Energy Resolution (FWHM) as a function of peaking time for strips A, C, E, and G of Sh0221, a GAPS flight detector after passivation with VTEC PI-1388 polyimide using final protocol (see Sec.~\ref{ss-detprep}). Measurements were performed at $-$37\,\textdegree{C} in the vacuum testing set-up outlined in Sec.~\ref{ss-noisetestingsetup}. \textit{Top Right:} Energy resolution (FWHM) at the optimal peaking time after passivation for each measured strip of the 12 eight-strip flight detectors, using optimized passivation procedure. \textit{Bottom Left:} Best-fit $A_{f}$ component (V\textsuperscript{2}) of the $1/f$ noise for the strips of these detectors. \textit{Bottom Right:} Best-fit leakage current (nA) for the strips of these detectors.}
\end{figure}

Detectors passivated with the final PI passivation protocol demonstrated good leakage current characteristics and indicated no degradation in X-ray energy resolution. The optimized passivation procedure was applied to a batch of 12 eight-strip flight detectors. Given time-constraints, these detectors were not tested for energy resolution before passivation, and not all strips were measured after passivation. At least four strips per detector were randomly sampled for post-passivation energy resolution measurements. Each measured detector strip had $\lesssim$\,4\,keV energy resolution, and was determined to have suitably low 1/f noise (see Fig.~\ref{f-passval}). A small fraction of detector strips demonstrated fitted leakage current above the acceptance criteria, but were deemed successful based on their energy resolution and 1/f noise.

\section{Effectiveness in Protecting Detectors\label{s-protection}}
\subsection{Motivation}
GAPS is scheduled for three LDB science flights. Therefore, the mission's success relies on passivated detector performance remaining stable for several years. Even with a passivation coating, precautions are taken to ensure that detectors have minimal exposure to humidity and organics. For long-term storage, detectors are stored in a vacuum sealed antistatic bag with desiccant in a freezer, which has been demonstrated to produce an environment with $<$\,5\% relative humidity (RH) at $-20$\,\textdegree{C}. During routine calibration and processing, lab spaces will be maintained at $<$\,30\% RH at room temperature. During integration, the modules will be purged continuously with N2 to mitigate humidity and organic outgassing.

\par Accelerated lifetime testing was conducted to assess the PI passivated detector's robustness to contamination from humidity (see Sec.~\ref{ss-accelhum}) and organic materials (see Sec.~\ref{ss-accelorg}) used in detector assemblies. Furthermore, a long-term detector performance monitoring program is ongoing to track detector performance over time (see Sec.~\ref{ss-longterm}).

\subsection{Accelerated Humidity Exposures\label{ss-accelhum}}
\subsubsection{Success Criteria}

\par Acceleration factors were computed by examining water vapor barrier penetration. The acceleration factor for humidity exposures is proportional to the number of water molecules hitting the detector surface at a given temperature and humidity, and can be expressed as~\cite{wayne1990accelerated}:
\begin{equation}\label{eq:humaccelfactor}
a=\frac{h P_w^{v}(T)}{h_0 P_w^{v}(T_0)}e^{-\frac{E_{p}}{R}(\frac{1}{T}-\frac{1}{T_0})}
\end{equation}
where $P_w^{v}(T)$ is the water saturation vapor pressure, $E_{p}$ is the activation energy for water diffusion into the polymer film, and $R$ is the gas constant. $T$ and $h$ are the temperature (Kelvin) and RH under test conditions, while $T_0$ and $h_0$ are the temperature (K) and RH during normal field use. $E_{p}$\,$\simeq$\,11\,kJ/mol was used to compute the humidity acceleration factor, based on an empirical measurement of water diffusion into polymer films~\cite{OgawaDiffusion}.

\par The success criteria for accelerated humidity exposures was motivated by the foreseen processing and storage conditions. To be deemed successful, a detector was required to undergo humidity exposures equivalent to 14~days at 50\%\,RH. This exposure is equivalent to $\sim$\,1~month of active work on a detector in typical lab conditions ($\sim$\,30\%\,RH at room temperature), and $\sim$\,20~years in a desiccated bag in a freezer at $\sim$\,--20\,\textdegree{C}.

\subsubsection{Experimental Set-up}
Accelerated humidity exposures were performed by placing a detector in an airtight chamber with a small water dish. The entire chamber was heated to 60\,\textdegree{C} for 3~hours, and the humidity in the chamber increased to $\sim$\,80\%\,RH. Each exposure is equivalent to $\sim$\,2~days at 23\,\textdegree{C} and 50\%\,RH. The chamber temperature and humidity was measured each minute using a logging hygrometer, and the temperature and humidity data was used to compute the acceleration factor and equivalent time of the exposure. 

\subsubsection{Results}
\begin{figure}[h]
\centering
\includegraphics[width=0.6\linewidth]{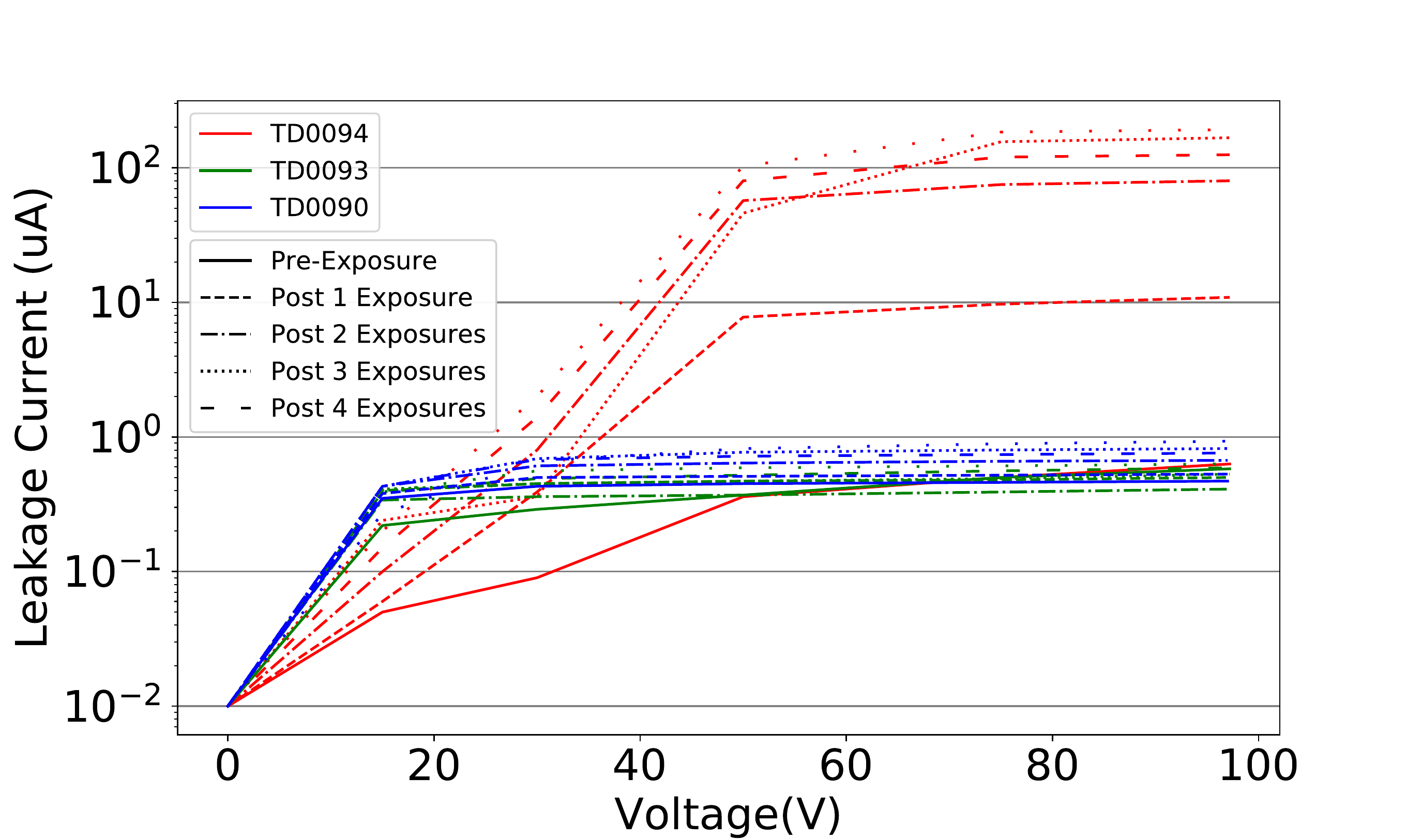}
  \label{fig:test3}
 \captionof{figure}{\label{f-humtesting}Room temperature leakage current measurement of TD0090 (passivated), TD0093 (passivated), and TD0094 (unpassivated) through successive humidity exposures. Solid lines indicate leakage current before exposure to humidity, whereas dashed and dotted lines indicate successive humidity exposures. In this test, the total accelerated exposure was comparable to 7~days at 50\%\,RH during field use, following Eq.~\eqref{eq:humaccelfactor}. The passivated detector performance was stable through all exposures, while the unpassivated detector leakage current degraded significantly.}
\end{figure}

Three detectors fabricated in-house were subjected to humidity exposures. Two detectors (TD0090 and TD0093) were passivated following the passivation protocol (see Sec.~\ref{sss-eresvalidation}) before being exposed to humidity. One detector (TD0094) was left unpassivated as control. Detector room temperature leakage current was measured before and after exposure to humidity. The passivated detectors demonstrated no degradation in leakage current, while the unpassivated detector's leakage current increased significantly (see Fig.~\ref{f-humtesting}).

Furthermore, a passivated 8-strip GAPS flight detector, Sh0070, was subjected to similar humidity testing and was resilient against degradation due to humidity. This detector's leakage current was stable through humidity exposures equivalent to 14~days at 50\%\,RH at room temperature.

\subsection{Accelerated Organics Exposures\label{ss-accelorg}}
\subsubsection{Success Criteria}
Once integrated, the detector modules consist of four eight-strip Si(Li) detectors, an ASIC board, and fluorosilicone and G10 detector retaining parts fastened into an aluminum frame. The modules are sealed on both sides with an aluminized polypropylene window. In this study, detectors were exposed to outgassing from organics from materials that are constituent in a detector module.

Assuming diffusion as the dominant source of outgassing, the outgassed material at a time $t$ is given by:

\begin{equation}\label{eq:diffdomoutgassing}
\Delta{m}(t,T)=f_{m} m_{0} \sqrt{\frac{t}{t_r}}e^{\frac{E_a}{R}(\frac{1}{T_r}-\frac{1}{T})}
\end{equation}

where $m_0$ is the initial mass of the organic, $E_a$ is the activation energy of the organic material, $R$ is the gas constant, and $f_{m}$ is the fractional mass loss at a reference time ($t_r$) and temperature ($T_r$)~\cite{NASAOutgassing}. Since we are mainly concerned with finding an acceleration factor, we note that the total mass loss (TML) $\propto t^{\frac{1}{2}}e^{-\frac{E_a}{RT}}$, where $T$ and $t$ are the accelerated exposure temperature and time during lab testing, and $T_0$ and $t_0$ are the temperature and time of exposure during normal field use. An acceleration factor for the lab exposures to organics can be found using the scaling relation:

\begin{equation}\label{eq:orgaccelfactor}
a \equiv \frac{t_0}{t} = e^{\frac{2E_a}{R}(\frac{1}{T_0}-\frac{1}{T})}
\end{equation}

In this study, detectors were exposed to a FR-4 circuit board, G10, fluorosilicone, and vacuum grease that will be used to install and seal the detector modules. The activation energies for these materials are not well measured, so to compute an acceleration factor, $E_{a}$\,=\,10\,kJ/mol was used, based on the typical activation energy for diffusion driven outgassing~\cite{NASAOutgassing}. To mitigate outgassing, the modules are equipped for N2 purge, however N2 purge is not always feasible. The success criteria was motivated by the approximate amount of time that detector surfaces will be exposed to organics when not being purged. Detectors were exposed to organics equivalent to $>$\,6 months of field use at room temperature, and were required to demonstrate no degradation in leakage current or energy resolution to be deemed successful.

\subsubsection{Experimental Set-up}
\par Accelerated organics exposures were performed by placing a detector in a chamber next to a hot plate, and heating the organic material on a hot plate to increase its TML. For each exposure, the organic material was heated to $\sim$\,70\,\textdegree{C} for 6~hours, an equivalent exposure of $\sim$\,30 days, using Eq.~\eqref{eq:orgaccelfactor}. Before and after each exposure, the detector's room temperature leakage current was measured. After achieving an equivalent exposure $>$\,6~months, the detector was sent to MIT for energy resolution testing.

\subsubsection{Testing}
\begin{figure}[h]
\centering
\begin{minipage}{.5\textwidth}
\centering
\includegraphics[width=\linewidth]{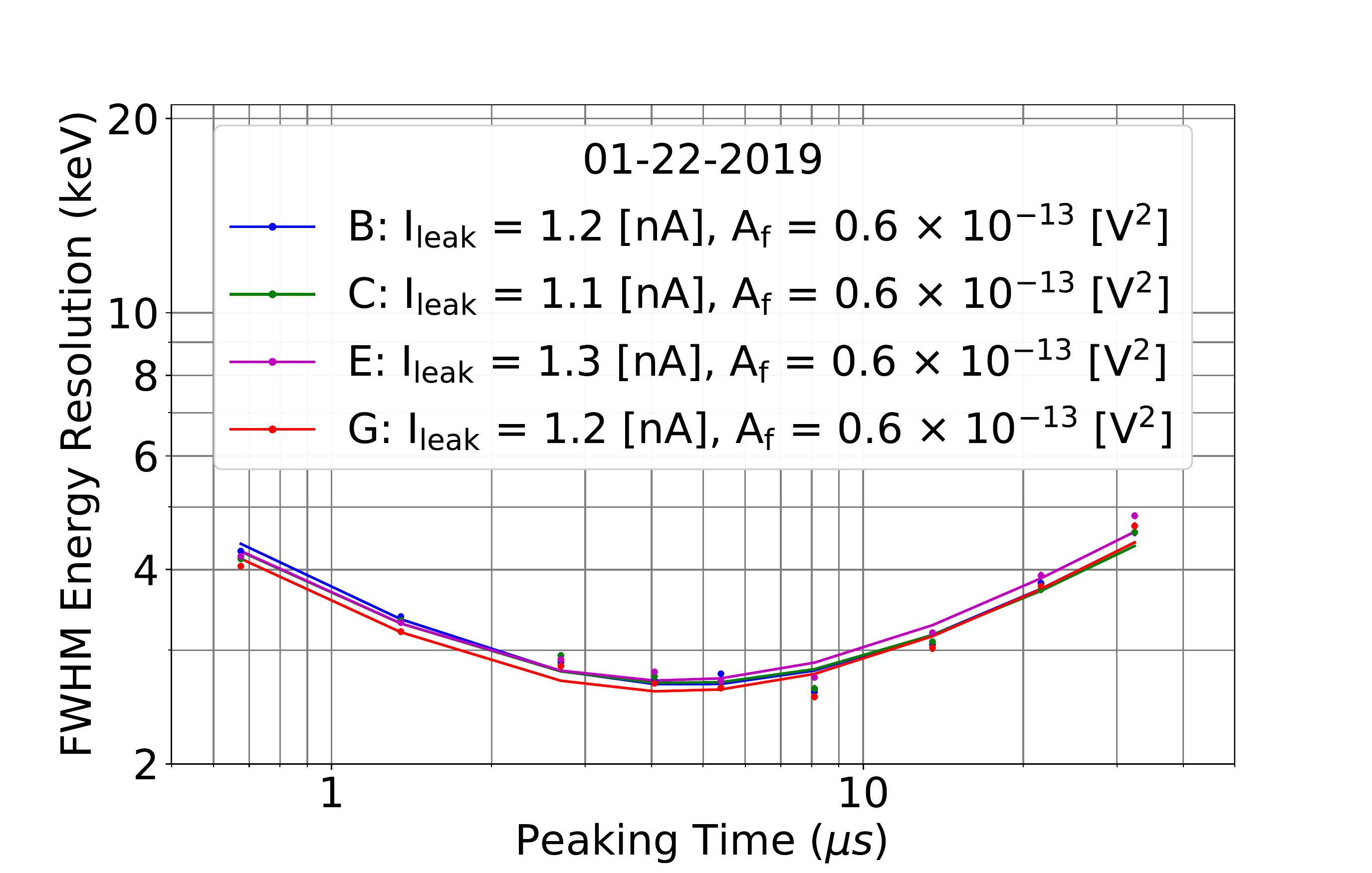}
\end{minipage}%
\begin{minipage}{.5\textwidth}
\centering
\includegraphics[width=\linewidth]{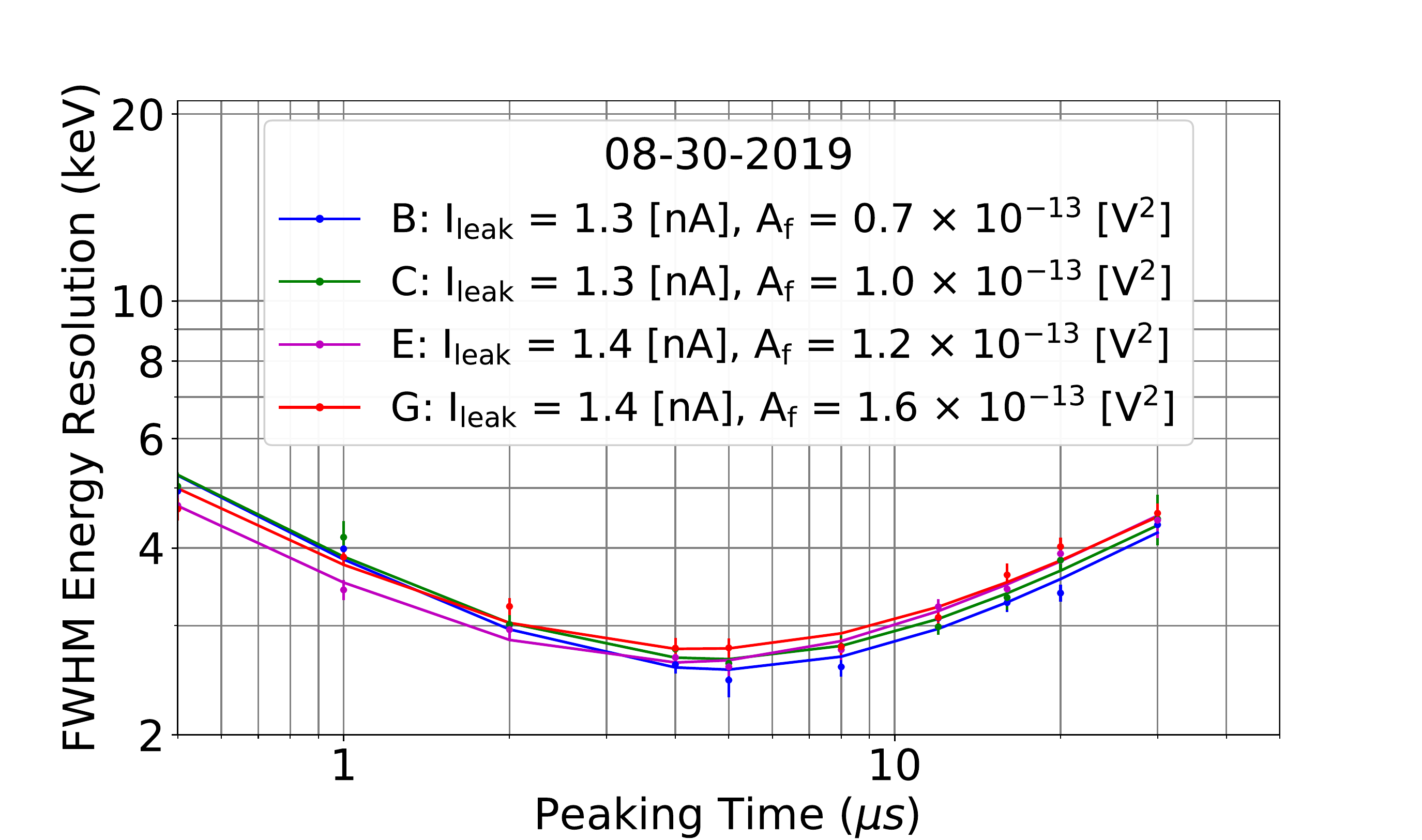}
\end{minipage}
\captionof{figure}{\label{f-orgtesting}\textit{Left:} Energy resolution (FWHM) as a function of peaking time for strips B, C, E, and G of Sh0079, an eight-strip flight detector. Measurements were performed at $\sim$\,--35\,\textdegree{C} immediately after passivation. \textit{Right:} Energy resolution (FWHM) as a function of peaking time for same strips of Sh0079, after organics exposures equivalent to 6 months of field use. No degradation in noise performance was observed, and the fit parameters of the noise model were consistent.}
\end{figure}

Two passivated 8-strip detectors, Sh0079 and Sh0161, were selected for exposure to organics. Both detectors demonstrated no change in room temperature leakage current characteristics through exposures to organics equivalent to $>$\,6~months field exposure to the materials in the GAPS detector modules at room temperature. After exposures, X-ray energy resolution was measured and no increase in energy resolution or 1/f noise was observed (see Fig.~\ref{f-orgtesting}).

\subsection{Long-Term Monitoring\label{ss-longterm}}

\begin{figure}[h]
\centering
\includegraphics[width=0.6\linewidth]{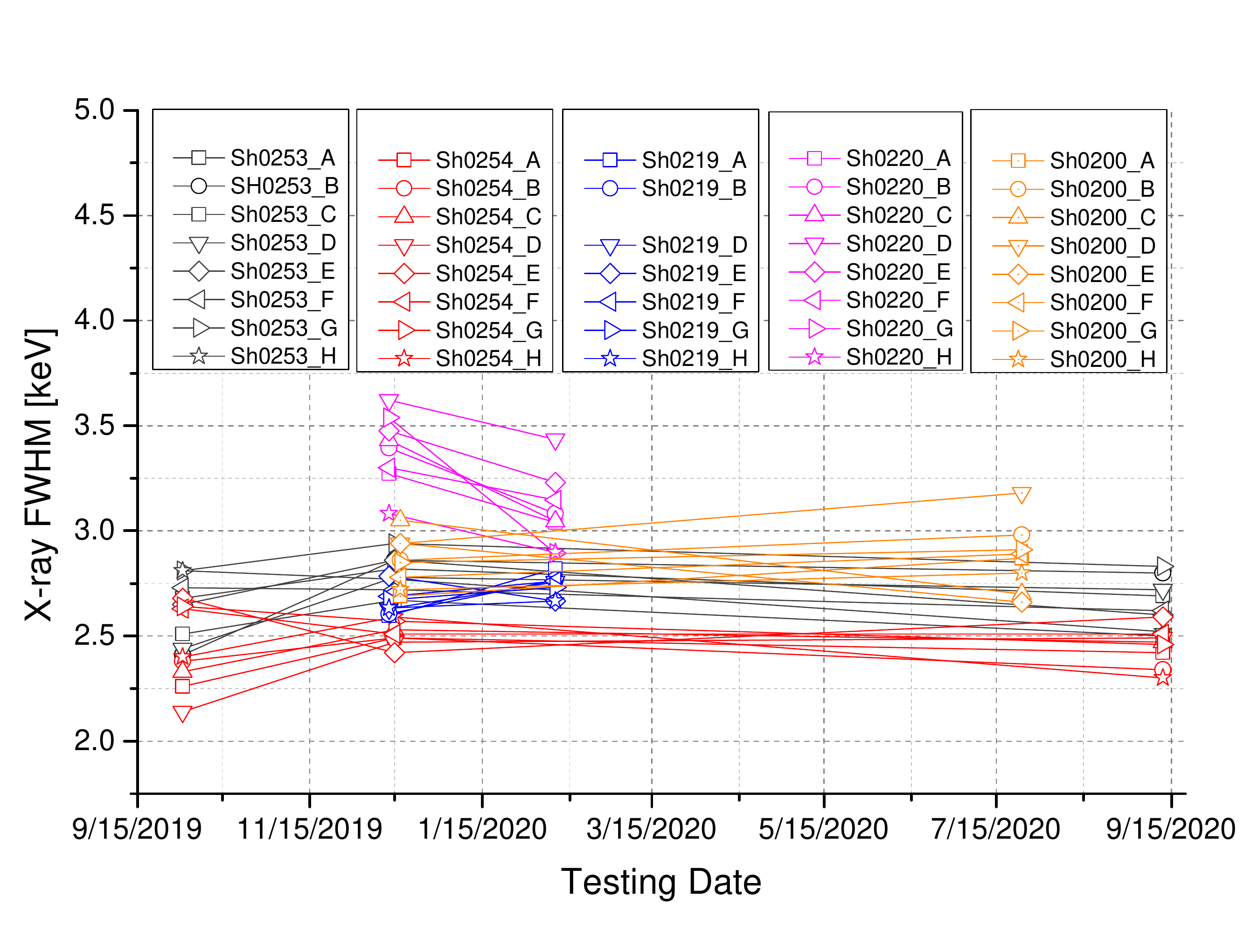}
\captionof{figure}{\label{f-longtermmonitor}Energy resolution (FWHM) at optimal peaking time for strips of five GAPS flight detectors, measured on different dates as part of long-term detector stability monitoring program. Passivated Si(Li) detectors demonstrate stable performance over year-long timescales.}
\end{figure}

A long-term monitoring program is ongoing to ensure detector stability. Over the course of a year, the energy resolution of passivated GAPS flight detectors was measured in the SUN chamber. Between measurements, detectors were stored in a dry box or in a vacuum-sealed bag with desiccant in a freezer at $\sim$--20\,\textdegree{C}. No detector degradation has been noted over the course of twelve months (see Fig.~\ref{f-longtermmonitor}).
\par In conjunction with accelerated lifetime testing, the preliminary results of long-term monitoring are promising. They indicate that storing the passivated GAPS flight detectors in a vacuum sealed bag with desiccant in a freezer at $\sim$\,--20\,\textdegree{C} is sufficient to maintain stable detector performance for the lifetime of the GAPS experiment. Moreover, it is not necessary to store these Si(Li) detectors under bias to mitigate Li re-distribution, which simplifies the long-term storage scheme needed to maintain detector performance. More details will be presented elsewhere.

\section{Conclusion\label{s-conclusions}}
We have demonstrated that Si(Li) surfaces can be successfully passivated for detectors operated at temperatures as high as $-35$\,\textdegree{C}. Polyimide and parylene-C were explored as passivation candidates. While parylene-C coatings demonstrated acceptable thermal and adhesion properties, excessive leakage current and poor deposition reproducibility made it unfit for large-scale passivation. After optimizing the application protocol, polyimide was selected as the passivation coating for GAPS Si(Li) detectors.
\par The PI passivated detector performance has been demonstrated to meet the requirement of $\leq$\,4 keV energy resolution (FWHM) in the 20--100 keV energy range in the GAPS operating temperature range ($-35$\,\textdegree{C} to $-45$\,\textdegree{C}). The passivation protocol is easy to apply by technicians, and the resulting passivation coating provides a robust barrier to humidity, organic, and particulate contamination. Passivated detector performance is stable over year-long time scales. This protocol is being employed to passivate 1440 flight detectors for the GAPS experiment, the first 491 of which have been passivated using the selected protocol. These passivated Si(Li) detectors will form the first large-area silicon detector array with X-ray spectral capabilities operated above cryogenic temperatures at high altitudes.

\section{Acknowledgements}
We thank SUMCO Corp.\ and Shimadzu Corp.\ for their cooperation in detector development. We also thank the GAPS collaboration for their consultation and support. This work is partially funded by the NASA APRA program (Grant Nos. NNX17AB44G and NNX17AB46G). K. Perez and M. Xiao receive support from the Heising-Simons Foundation. F. Rogers is supported by the NSF Graduate Research Fellowship (Grant No. 1122374). This work is partly supported in Japan by JAXA/ISAS Small Science Program FY2017. H. Fuke receives support from JSPS KAKENHI grants JP26707015, JP17H01136, and JP19H05198 and from the Mitsubishi Foundation. M. Kozai receives support from JSPS KAKENHI grants JP17K14313 and JP20K14505. Y. Shimizu receives support from JSPS KAKENHI grant JP20K04002 and Sumitomo Foundation grant.
\section*{References}
\footnotesize
\bibliography{referenceswLinksv2}

\end{document}